\documentclass[aps,prb,notitlepage,superscriptaddress,reprint]{revtex4-1}
\pdfoutput=1
\usepackage{amsfonts}
\usepackage{amsmath}
\usepackage{amssymb}
\usepackage{graphicx}
\usepackage{bm}
\usepackage{color,soul}
\usepackage{float}
\usepackage{indentfirst}
\usepackage{comment}

\begin{document}
\title{Pseudomagnetic fields in fully relaxed twisted bilayer and trilayer graphene}

\author{A. Ceferino}
\email{adrian.ceferino@imdea.org}
\address{Imdea Nanociencia, Faraday 9, 28015 Madrid, Spain}

\author{F. Guinea}
\address{Imdea Nanociencia, Faraday 9, 28015 Madrid, Spain}
\address{Donostia International Physics Center, Paseo Manuel de Lardizabal 4, 20018 San Sebastian, Spain}
\begin{abstract}
We present simple models to describe the in-plane and the out-of-plane lattice relaxation in twisted bilayer and symmetrically twisted trilayer graphene. Analytical results and series expansions show that for twist angles $\theta>1$º, the in-plane atomic displacements lead to pseudomagnetic fields weakly dependent on $\theta$. In symmetrically twisted trilayer graphene, the central layer in-plane relaxation is greatly enhanced. The joint effect of the relaxation-induced pseudoscalar potentials and the associated energy difference between interlayer dimer and non-dimer pairs resulted in a significant electron-hole asymmetry both in twisted bilayer and trilayer graphene.



\end{abstract}
\maketitle
\section{Introduction}

\indent Twisted bilayer graphene shows a very complex electronic phase diagram for small twist angles arising from the existence of narrow electronic bands for certain angles referred to as ``the magic" angles\cite{Cetal18a,Cetal18b,HLSCB19,KKTV19,ELFS22,MagicAnglesBerevig,magicring}. The width and shape of the electronic bands near these magic angles are highly sensitive to many perturbations, such as strains\cite{BYF19,Metal21}, interactions\cite{GW18}, or the effect of the substrate\cite{CPG21,SZM21,SPCSJ21,MS21,LSN21}. These effects typically increase the width of the narrow middle bands, (whose width varies between 4 and 40 meV), and can also give rise to small gaps between the valence and conduction bands. \\
\indent The interaction between the two twisted layers in tBLG leads to lattice relaxation\cite{NK17,GW19,KN20,CACCNL20,Leconte,YazyevRelax,VanWijk,O19,xie2023lattice,abinitio}, which shrinks the highly unstable AA regions while enlarging the more energetically favorable AB regions effectively reducing the twist angle in the AB regions while increasing the local twist angle in the AA-stacking zones. Lattice relaxation has been shown to play a pivotal role both in the piezoelectric and ferroelectric properties of twisted TMDs\cite{weston2020atomic,enaldiev2020stacking,vovaFerro,weston2022ferro} as well as in the charge-density-wave phases of twisted NbSe$_{2}$\cite{mchughNb}. In graphitic systems, atomic relaxation has been shown to determine the out-of-plane polarisation of twin boundaries\cite{Aitorferro} while in supermoir\'e systems such as the non-symmetric twisted trilayer graphene, lattice relaxation leads to a clear separation between the flat band and the highly dispersive Dirac cone\cite{PCWTJ21,Hetal21,STS22,Wurelaxtrilayer,Koshinotrilayerarxiv,SamajdarTrilayer,Lintrilayer,LadoBandflat,McDonaldMirror,Kaxirastrilayer,Popovtrilayer,Kimtrilayers,Jungtrilyer,PhongTrilayer,Bernevigtrilayer,Moratrilayer,CraigSTM,CoryDeantrilayer}. For sufficiently small twist angles, $\theta \lesssim 0.5^\circ$, a domain pattern, made of $AB$ and $BA$ regions separated by domain walls emerges, leading to a conducting network of topologically protected 1D channels\cite{TNK20,SGG12,VortexKast,ChannelsRecther,Efimkin,walet1D}. Near the first magic angle, where the Fermi velocity vanishes\cite{BM11,TKV19}, $\theta \approx 1^\circ$, the in-plane atomic displacements lead to an effective magnetic field\cite{NK17,VKG10,Setal20} with opposing sign in the two layers and two valleys which heavily renormalizes the intervalley hopping.\\ 
\indent In this manuscript, we analytically obtain the lattice relaxation-induced pseudomagnetic field\cite{VafekRelax,VafekCont} profile for twist angles near the first magic angle as well as the scalar potential derived from the in-plane and out-of-plane lattice relaxation. Then, we added to our continuum tBLG and tTLG Hamiltonian an additional term accounting for the energy difference between the interlayer $p_{z}$ orbital dimer and non-dimer pairs\cite{FullSWC}.
\begin{figure}[t!]
    \centering
    \includegraphics[width =3.6in, height=4.8in]{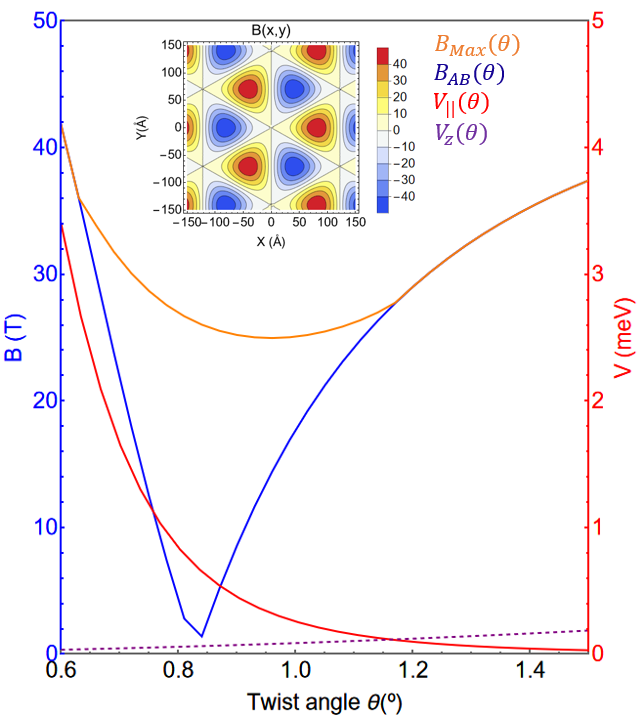}
    \caption{\footnotesize Relaxation-induced pseudomagnetic and pseudoscalar potentials as a function of twist angle $\theta$ in twisted bilayer graphene. The two pseudoscalar potentials $V_{z}$ and $V_{||}$ come from the out-of-plane and in-plane strains respectively. (Inset) Pseudomagnetic field profile in twisted bilayer graphene obtained via the linear response formalism presented in Eq.(\ref{eqn:NonlinU}) and (\ref{eqn:NonlinUy}) for a twist angle of $1^\circ$. In this formalism, the pseudomagnetic fields localized in the AB and BA region are independent of the twist angle and the pseudoscalar potential resulting from the adhesion forces is always equal to zero.}
\label{fig:PotandB}
\end{figure}
We find that the in-plane deformations not only shift the position of the tBLG magic angle towards lower twist angles but also form an entire range of twist angles $\theta\in[0.8-0.85]$ where both the lowest conduction and highest valence bands are very flat\cite{carr2019exact}. Similarly, in a tTLG made of three AAA-stacked layers with the middle layer twisted at an angle $\theta$ relative to the top and bottom graphene membranes (see Fig.\ref{fig:Setup} b), the collapse of the Fermi velocity towards lower angles and the formation of a magic angle plateau was found even more pronounced. Finally, lattice relaxation hugely altered the electron-hole asymmetry both in twisted bilayer and trilayer graphene while it preserved the topological protection of the Dirac cones.

\section{Lattice relaxation in \MakeLowercase{t}BLG and \MakeLowercase{t}TLG}\label{section:geom}

\begin{figure}[t!]
    \centering
    \includegraphics[width =3.6in, height=6.6in]{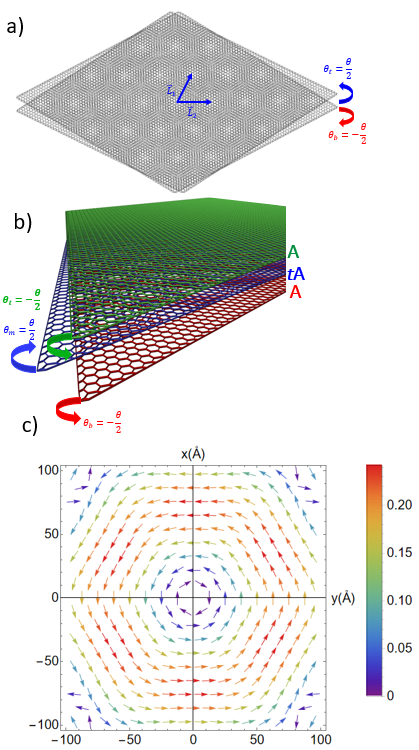}
    \caption{\footnotesize a) Top view of tBLG for an arbitrary twist angle $\theta$. b) Twisted trilayer graphene setup under consideration with a middle layer twisted at an angle $\theta$ relative to its top and bottom counterparts. c) Characteristic shape of the atomic displacement field of the top layer in units of \AA\quad  when a twist angle $\theta$ is applied.}
\label{fig:Setup}
\end{figure}

Lattice relaxation originates from the competition between van der Waals forces which tend to enlarge the $AB/BA$ regions and the graphene in-plane stiffness which counteracts such tendency. We start our analysis considering the superlattice geometry presented in Fig.\ref{fig:Setup}. Firstly we orient the monolayer graphene such that their lattice vectors are taken as $\bold{a}_{1}=a\Big(1,0\Big),\bold{a}_{2}=a\Big(\frac{1}{2},\frac{\sqrt{3}}{2}\Big)$ where $a=2.46$\AA, which prescribes the following two reciprocal lattice vectors of the monolayer graphene membrane, $\bold{g}_{1}=\frac{2\pi}{a}\Big(0,-1\Big),\bold{g}_{2}=\frac{2\pi}{a}\Big (\frac{\sqrt{3}}{2},-\frac{1}{2}\Big)$. Considering an AA-stacked bilayer with lattice vectors $\bold{a}_{1},\bold{a}_{2}$ we first apply a $\theta/2$ counter-clockwise rotation of the top layer around the center of the hexagonal lattice. Then we rotate the bottom layer by an angle $-\theta/2$ relative to the same rotation axis, shifting the two originally overlayed carbon atoms located at $\bold{r}_{0}$ relative to the rotation center by an amount $\Delta\bold{r}=\bold{r}_{t}-\bold{r}_{b}=\theta \hat{z}\times \bold{r}_{0}$.
This new geometrical arrangement results in an approximately periodic pattern (see Fig.\ref{fig:Setup} a)\cite{} characterized by a new periodic Moir\'e lengthscale defined as, 
\begin{align}
    L_{M}=\frac{a}{2\sin{\Big(\theta/2\Big)}}.\label{eqn:Mlength}
\end{align}
Given our previously defined monolayer orientation, the Moir\'e reciprocal superlattice vectors are obtained from the difference in wavevector between the rotated graphene reciprocal lattice vectors,
\begin{align}\label{eqn:Gdef}       
\bold{G}_{i}=R(-\theta/2)\bold{g}_{i}-R(\theta/2)\bold{g}_{i}=\bold{g}^{b}_{i}-\bold{g}^{t}_{i}=-\theta\hat{\bold{z}}\times \bold{g}_{i}.
\end{align}
Therefore, $\bold{G}_{1}=\frac{2\pi}{L_{M}}\Big(-1,0\Big)$ and $\bold{G}_{2}=\frac{2\pi}{L_{M}}\Big(-\frac{\sqrt{3}}{2},-\frac{1}{2}\Big)$. We model the adhesion van der Waals potential $W(\vec{r})$ between the displaced layers by Fourier expanding the potential in Moir\'e reciprocal lattice vectors as presented in Ref.\onlinecite{CWTCLK18,Kaxirasbil}. Considering the lowest and more dominant harmonics, we obtain the following approximate expression for the adhesion potential which satisfies all the symmetries of the Moir\'e supercell\cite{NK17,GW19},
\begin{align}
    W(\vec{r})\approx V_{vdW} \sum_{i= 1, 2, 3} \cos \left[ \vec{g}_i ( \vec{r}_t - \vec{r}_b ) \right], 
    \label{eqn:Wpot}
\end{align}
where the three $\vec{g}_i$ are three trigonally symmetric reciprocal lattice vectors of each graphene layer and $\vec{r}_t,\vec{r}_b$ label the spatial coordinates within the $xy$-plane of a point in the top and bottom layer respectively. The amplitude $V_{vdW}$ is set to $V_{vdW}=0.9$meV\AA$^{-2}$, following the electron-phason study presented in Ref.\onlinecite{O19}. Due to the interplay between the adhesion and the elastic forces, the top and bottom graphene membrane experience an atomic displacement field $\vec{u}^{t}$ and $\vec{u}^{b}$ respectively. To balance the force that the top layer does onto the bottom one and vice-versa, we assume that lattice relaxation induces at each point of the two graphene membranes a perfectly antisymmetric displacement field such that $\vec{u}_{t}=-\vec{u}_{b}=\vec{u}$. Therefore, each atomic location of the top and bottom layer after twisting and relaxing becomes 
\begin{align}
\vec{r}_t &= R_{\theta/2} ( \vec{r} )+\vec{u} \nonumber \\
\vec{r}_b &= R_{- \theta/2} ( \vec{r} )-\vec{u},\label{eqn:Rrel} 
\end{align}
where $R_{\theta/2}$ is the rotation operator by a twist angle of $\theta/2$. Eq.(\ref{eqn:Wpot}) can be rewritten in terms of the supercell reciprocal lattice vectors as,
\begin{align}
    \vec{g}_i \left[ R_{\theta/2} ( \vec{r} ) - R_{- \theta/2} ( \vec{r} ) \right] \approx \vec{G}_i \vec{r},
\end{align}
which leads to
\begin{align}
    W ( \vec{r} ) &\approx V_{vdW} \sum_{i=1,2,3} \cos ( \vec{G}_i \vec{r} + 2 \vec{g}_i \vec{u} ).
    \label{eqn:vdW2}
\end{align}
Assuming that $V_{vdW}>0$, the function $W(\vec{r})$ has a maximum ($W_{max}=3V_{vdW}$) at the center of the real space unit cell ($AA$ region), and two minima, $W_{min}=-(3V_{vdW})/2$ located at the corners of the Moir\'e unit cell\cite{KolmoCrespi} ($AB$ and $BA$ regions). Note that in Eq.(\ref{eqn:vdW2}), the reciprocal lattice vectors $\vec{G}_i$ and $\vec{g}_i$ are orthogonal with magnitude $|\vec{G}_i|=(4\pi)/(\sqrt{3}L)$ and $|\vec{g}_i|=(4\pi )/(\sqrt{3}a)$ respectively, where $L$ and $a$ are the lattice constants of the Moir\'e supercell and of each isolated monolayer. The van der Waals forces on the atoms experienced at the top and bottom layers can be obtained by taking the gradient of the potential in Eq.(\ref{eqn:vdW2}) with respect to $\pm\vec{u}$. Assuming that $\vec{G}_{i}\vec{r}>>2\vec{g}_{i}\vec{u}$, the force of the top and bottom layer become,
\begin{align}
   \vec{F}_{t,b}(\vec{r})&=\mp V_{vdW}\sum \vec{g}_i \sin(\vec{G}_i\vec{r}).
   \label{eq:forces}
\end{align}
These forces define a transverse vector field normal to the van der Waals potential which induces a static atomic displacement field towards their equilibrium position. By balancing the force in Eq.(\ref{eq:forces}) with that of a frozen transverse acoustic phonon whose elastic energy is only given by the Lamé coefficient $\mu=9.25$eV\AA$^{-2}$[\onlinecite{ZKF09}], we obtain an equilibrium condition for the atomic displacement at each point of the two graphene membranes,
\begin{align}
 u_{t}=-\vec{u}_{b} &\approx  \frac{V_{vdW}}{\mu G^2}   \sum_{i=1,2,3} \vec{g}_i \sin ( \vec{G}_i \vec{r} ).
 \label{eqn:Ulin}
\end{align}
\begin{widetext}
\begin{figure*}[t!]
    \centering
    \includegraphics[width =7.2in, height=4in]{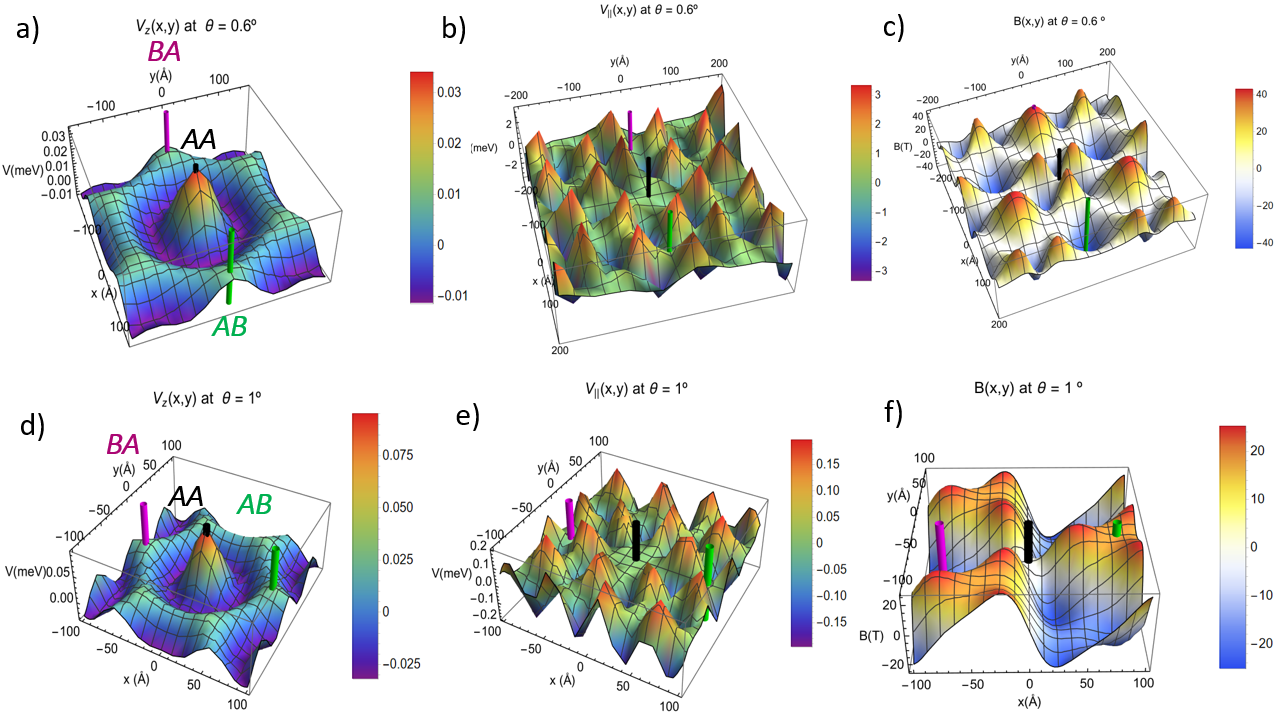}
    \caption{\footnotesize Figures a) and d) show the pseudoscalar potential profile at $\theta=1^\circ$ and $\theta=0.6^\circ$ due to the different in interlayer distances in the Moir\'e unit cell in units of meV. Figures b) and e) show the pseudoscalar potential profile at $\theta=1^\circ$ and $\theta=0.6^\circ$ due to the in-plane atomic displacements resulting from the interlayer adhesion forces in units of meV. Figures c) and f) show the pseudomagnetic field distribution at twist angles $\theta=1^\circ$ and $\theta=0.6^\circ$ in units of Teslas. The strain field used for these plots was the strain field of the top layer $\vec{u}^{t}$.}
\label{fig:Pseudoprofiles}
\end{figure*}
\end{widetext}
From the above atomic displacement fields, we predict a vanishing deformation potential\cite{Ando} ($V=\Upsilon(u_{xx}+u_{yy})$ where $\Upsilon=4.5$eV[\onlinecite{Upsilonparam}] is the work function of graphene) and a twist-angle independent pseudomagnetic field\cite{guineapseudo,lowpseudo} given by $\vec{B}=\vec{\triangledown}\times\vec{A}$ where $\vec{A}(\vec{r})=\pm\frac{3}{4}\frac{t|\beta|}{v_F}(u_{xx}-u_{yy},-u_{xy}-u_{yx})$,
\begin{align}
    B (\vec{r}) = \mp \sqrt{3} \pi \frac{t \beta}{v_F} \frac{V_{vdW}}{\mu d} \sum_{i=1,2,3} \sin ( \vec{G}_i \vec{r} ).
    \label{eqn:Blin}
\end{align}
In the above expression, $t\approx 2.7$eV[\onlinecite{AccurateTB}] is the hopping parameter between nearest neighboring carbon-atom $p_z$ orbitals and $\beta$ is the Gr\"{u}nesein parameter[\onlinecite{Raman}] $\beta\equiv-\frac{\partial ln(t)}{\partial ln(a_{c})}\approx 3$, with $a_{c}=a/\sqrt{3}\approx 1.42$ \AA \, being the distance between nearest neighboring atoms. The parameter $v_F=\frac{\gamma_{0}\sqrt{3}a}{2\hbar}\approx 5.59$ eV\AA \, is the monolayer graphene Fermi velocity. From Eq.(\ref{eqn:Blin}), the characteristic magnetic length of electrons located at the magnetic hot spots in the AB and BA is obtained as,  
\begin{align}\label{eqn:lBmag}
    \ell_M =a\sqrt{\frac{1 }{3 \pi} \frac{1}{| \beta |}\frac{\mu}{V_{vdW}}}\approx 7.8nm.
\end{align}
for $V_{vdW}/\mu \approx 10^4$. This pseudomagnetic field averages to zero over the unit cell, and reaches $\approx \pm 42$ T at the $AB$ and $BA$ corners (see inset in Fig.(\ref{fig:PotandB})), in reasonable agreement with the results presented in Ref.\onlinecite{NK17} and \onlinecite{Setal20}. To improve this simple formulation of the in-plane relaxation in tBLG, we plug the obtained $(u_{x},u_{y})$ from Eq.(\ref{eqn:Ulin}) into the adhesion term in Eq.(\ref{eqn:vdW2}) and solve the Euler-Lagrange equations via the Fourier transform method (see Appendix.\ref{app:elasticity}) to recalculate a new atomic displacement field $(\delta U_{x},\delta U_{y})$. The solutions of the resulting strain field $(\delta U_{x},\delta U_{y})$ have the form
\begin{table} [b!]
	\begin{tabular}{|c|c|c|c|}
 		\hline\hline
		$  \omega_{AA} $ & $ 79.7 $~meV &
		$ \omega_{AB} $& $ 97.5 $~meV \\
		$ \gamma_{0} $& $ 2.6$~eV &
		$ \gamma_{2} $& $ -17 $ ~meV \\  
		$ \Delta $& $25$ meV & 
	    $  v_{F} $& $ 5.59 $~eV\AA \\
		\hline
		$ \mu $ & $ 9.57$ ~eV\AA$^{-2}$ &
	    $ \lambda $ & $ 3.25$~eV\AA$^{-2}$ \\ 
     	$ \Upsilon$ & $4.5$~eV &
        $ V_{VdW} $ & $0.9$ meV\AA$^{-2}$ \\
        $ \beta $ & $3 $ &
        $ a_{cc}$ & $1.42$ \AA \\ 
        $ a $& $2.46 $ \AA & $ d $& $ 3.35  $ \AA \\ 
        $ \Delta h $& $0.24 $ \AA & $  $& $  $ \\ \hline
\end{tabular}\label{tbl:param}
\caption{\footnotesize (Top) Parameters used for the calculation of the electronic bandstructure in twisted bilayer and trilayer graphene. (Bottom) Parameters used for the calculation of the elastic deformations in the two membranes.}
\end{table}
\begin{widetext}
\begin{align}\label{eqn:NonlinU}
    &
    \delta U_{x}(\vec{r})=\frac{\sum^{3,\infty\infty,\infty}_{i=1,l=0,m=0,n=0}((2\mu+\lambda)G_{l,m,n,iy}^{2}+\mu G_{l,m,n,ix}^{2})A^{\pm l,\pm m,\pm n,i}_{x}\sin{((\pm l\vec{G}_{1}\pm m\vec{G}_{2}\pm n\vec{G}_{3}+\vec{G}_{i})\vec{r})}}{\mu(2\mu+\lambda)G_{l,m,n,i}^{4}}\\ \nonumber
    &
    -\frac{\sum^{3,\infty\infty,\infty}_{i'=1,l'=0,m'=0,n'=0}(\mu+\lambda)G_{l',m',n',i'y}G_{l',m',n',i'x}A^{\pm l',\pm m',\pm n',i'}_{y}\sin{(\pm l'\vec{G}_{1}\vec{r}\pm m'\vec{G}_{2}\vec{r}\pm n'\vec{G}_{3}\vec{r}+\vec{G}_{i'}\vec{r})}}{\mu(2\mu+\lambda)G_{l',m',n',i'}^{4}}, \nonumber
\end{align}
\end{widetext}
with $\delta U_{y}(\vec{r})$ obtained by replacing $G_{x}$ by $G_{y}$ and $G_{y}$ by $G_{x}$ into Eq.(\ref{eqn:NonlinU}),
\begin{equation}\label{eqn:NonlinUy}
    \delta U_{y}(\vec{r})=\delta U_{x}(\vec{r})[G_{x}\longrightarrow G_{y},G_{y}\longrightarrow G_{ix}].\\ 
\end{equation}
In Eq.(\ref{eqn:NonlinU}), the terms $A^{\pm l,\pm m,\pm n,i}_{x}$ and $A^{\pm l,\pm m,\pm n,i}_{y}$  are defined as in Eq.(\ref{eqn:Adeff}) and (\ref{eqn:Adefth}), of Appendix.\ref{app:elasticity}, while $\mu=9.57$eV\AA$^{-2}$,$\lambda=3.24$eV\AA$^{-2}$[\onlinecite{ZKF09}] are the graphene's Lamé and shear modulus coefficients respectively. Finally, the terms $G_{l,m,n,i}$ are defined as $G_{l,m,n,i}=l\vec{G}_{1}+m\vec{G}_{2}+ n\vec{G}_{3}+\vec{G}_{i}$ where $\vec{G}_{1}=\frac{2\pi}{L_{M}}(1,0)$, $\vec{G}_{2}=R(2\pi/3)\vec{G}_{1}$ and $\vec{G}_{3}=R(2\pi/3)\vec{G}_{2}$ . As shown in the comparison presented in Fig.\ref{fig:comparison} between our formalism and the numerical method developed in Ref.\onlinecite{NK17}, the description provided in Eq.(\ref{eqn:NonlinU}) and (\ref{eqn:NonlinUy}), correctly catches both the pseudomagnetic and pseudoscalar potential down to angles $\theta>0.5$º where domain walls start forming. \\
\indent While Eq.(\ref{eqn:NonlinU}) and (\ref{eqn:NonlinUy}) accurately model the in-plane relaxation of two coupled membranes, a realistic model of atomic relaxation in tBLG requires taking into account the out-of-plane deformation \cite{Canteleoutofplance,LinPressure,Leconte,UchidaAtomic} of its constituent membranes. Given its limited coupling with the planar strain field (see Fig.\ref{fig:PotandB}), we will account for the effect of out-of-plane atomic displacements by simply plugging into the z-component of the strain field tensor a height profile of the form
\begin{align}\label{eqn:hprof}
h(\Vec{r})=\Big(d+\frac{3h_{0}}{2}\Big)+h_{0}\sum^{i=3}_{i=1}\cos{\Big(\Vec{G}_{i}\bold{\cdot}\Vec{r}\Big)},
\end{align}
where $h_{0}=\frac{\Delta h}{4}= 0.06$\AA [\onlinecite{CACCNL20}] is a quarter of the difference in the interlayer separation between the AA and AB-stacked regions, and $d=3.35$\AA is the interlayer distance in Bernal bilayer graphene. As shown in Appendix.\ref{app:elasticity}, this additional term adds to the solutions of the Euler-Lagrange equations $(\delta U_{x},\delta U_{y})$ a radially symmetric term $(\delta U^{h}_{x},\delta U^{h}_{x})\propto|G|h^{2}_{0}\sin{(\vec{G}_{i}+\vec{G}_{j})}$ which adds $\sim 0.1-0.2$meVs to the pseudoscalar potential $V(\vec{r})\propto (u_{xx}+u_{yy})$, becoming the main contributor to the pseudoscalar potential only at angles $\theta>1$º (see Fig.\ref{fig:PotandB}).

\section{Continuum model of fully relaxed \MakeLowercase{t}BLG and \MakeLowercase{t}TLG}\label{section:cont}

\begin{figure}[b!]
    \centering
\includegraphics[width =3.5in, height=7.4in]{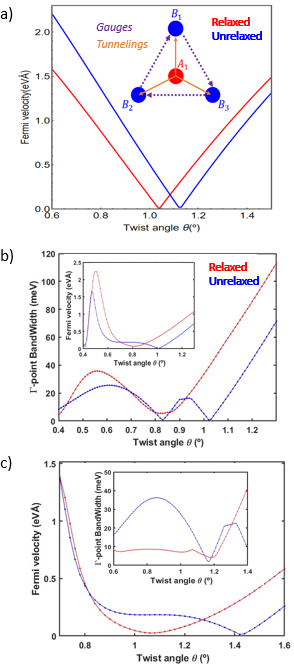}
    \caption{\footnotesize a) Renormalised Dirac-point band velocities of the tBLG as a function of twist angle $\theta$ with and without relaxation. The velocities were calculated using the two-band model described in Appendix.\ref{app:twoband}and the linear response formalism in Eq.\ref{eqn:Ulin}. (Inset) Schematic view of the tripod Hamiltonian. The orange lines depict the interlayer tunneling terms while the purple ones show the connections due to the pseudogauge fields. b) $\Gamma$-point bandwidth and Fermi velocity in tBLG for the unrelaxed and relaxed case. c) Fermi velocity and $\Gamma$-point bandwidth in tTLG both for the relaxed and the unrelaxed case.}
\label{fig:magic}
\end{figure}

\indent To compute the electronic bandstructure of tBLG and tTLG with atomic relaxation, we will tune the commonly used Bistrizer and MacDonald (BM) model\cite{BM11} to accommodate all the effects derived from atomic relaxation. In its simplest form, the BM model consists of two rotated Dirac Hamiltonians connected by a unique interlayer hopping parameter\cite{,LPN07} $\omega$ which couples any two Dirac cones in opposite layer separated an amount $\Delta K=k_{\theta}=2K\sin(\theta/2)$ in momentum space\cite{BernevigPT}. Written in the four-basis $\Psi^{t}_{A},\Psi^{t}_{B},\Psi^{b}_{A},\Psi^{b}_{B}$, the BM Hamiltonian has the form,
\begin{align}\label{eqn:BM}
\mathcal{H}_{BM}=\begin{pmatrix}
    R_{\theta/2}[-v_{F}i\Vec{\sigma}\Vec{\triangledown}]& \mathcal{T}\\
    \mathcal{T}^{\dagger} & R_{-\theta/2}[-v_{F}i\Vec{\sigma}\Vec{\triangledown}] \\ \nonumber
\end{pmatrix},\\
\mathcal{T}=\sum^{j=2}_{j=0}\omega\begin{pmatrix}
    1& e^{ij\phi}\\
    e^{-ij\phi}&1 \\
\end{pmatrix}e^{-i\Delta\vec{\bold{K}}_{j}\vec{r}},
\end{align}
where $R_{\theta/2}$ is a rotation operator twisting the Dirac Hamiltonian an angle $\theta/2$ and $\Delta \vec{\bold{K}}_{1}=k_{\theta}(0,-1),\Delta \vec{\bold{K}}_{2}=k_{\theta}(\frac{\sqrt{3}}{2},1/2),$ and $\Delta \vec{\bold{K}}_{3}=k_{\theta}(-\frac{\sqrt{3}}{2},1/2)$ are the momentum differences between neighbouring Dirac cones. Due to the different interlayer distance between the AA and AB-stacking patches, the interlayer hopping parameter in the AA regions ($\omega_{AA}$) and the AB or BA regions ($\omega_{AB}$) will be slightly different. To account for this difference, we will use the parameterization presented in Ref.\onlinecite{KoshWann} for tBLG, which obtained $\omega_{AA}$ and $\omega_{AB}$ by Fourier transforming the Slater-Koster interlayer tunneling amplitude of an AA and AB stacked-bilayer around the $K$-point. Therefore, the interlayer tunneling becomes
\begin{align}
    \mathcal{T}=\sum^{j=2}_{j=0}\begin{pmatrix}
    \omega_{AA}& \omega_{AB}e^{ij\phi}\\
    \omega_{AB}e^{-ij\phi}& \omega_{AA} \\
\end{pmatrix}e^{i\Delta\vec{\bold{K}}_{j}\vec{r}},
\end{align}
On top of the interlayer hopping renormalisation due to the difference in interlayer distance, it is necessary to consider to effect of in-plane relaxation in $\omega_{AA}$ and $\omega_{AB}$. To account for such an effect, we start from the Fourier expansion of the interlayer tunneling term in Eq.(\ref{eqn:BM}),
 and, as in Ref.[\onlinecite{VafekRelax}] we replace $\vec{r}_{t/b}$ with $\vec{r}_{t/b}\longrightarrow \vec{r}_{t/b}+\vec{u}_{t/b}$. Plugging the expression derived in Eq.(\ref{eqn:Ulin}) and Taylor expanding the complex exponential in Eq.(\ref{eqn:BM}), we can obtain a first-order correction to the interlayer tunneling terms $\omega_{AA}$ and $\omega_{AB}$ (see Appendix.\ref{app:continuumapend}. for a more detailed discussion) whose order of magnitude can be estimated as, 
\begin{align}
\frac{\delta \omega_{AA/AB}}{\omega_{AA/AB}} \approx i\frac{2V_{vdW}}{\sqrt{3}\mu}\frac{L_{M}}{a}\approx \frac{0.042}{\theta}
\end{align}
Since $\delta w_{AA}\ll\omega_{AA}$ and $\delta w_{AB}\ll\omega_{AB}$, we will neglect the effect of in-plane relaxation in the interlayer hopping amplitudes. \\
\indent To better capture the effect of atomic relaxation within the continuum model, we need to account for the energy difference in each carbon $A/B$ site due to its surrounding atomic registry. In an AB bilayer graphene structure, it is well known that the $A^{b}-B^{t}$ dimer pairs are at higher energy compared to their non-dimer $A^{t}-B^{b}$ counterparts. Therefore, in systems such as tBLG, this energy difference labeled by the parameter $\Delta=25$meV[\onlinecite{yingraphite}] is expected to create a difference in chemical potential between the AB/BA stacked regions and the more energetically unfavorable AA regions. As shown in Ref.[\onlinecite{FullSWC}], this additional term can be introduced in the tBLG continuum model by projecting the crystal potential of one layer into the other one, leading to an additional term $\mathcal{E}^{b}$ for its bottom layer,
\begin{widetext}
\begin{align}
\mathcal{E}^{b}=\begin{pmatrix}
\mathcal{E}^{bA} & 0\\
0 & \mathcal{E}^{bB}
\end{pmatrix}=\frac{2\Delta}{9}\sum^{i=2}_{i=0}\begin{pmatrix}
\cos{(\bold{G}_{i}\bold{r})}(1+\cos{(\bold{g_{i}}\bold{\tau_{b}})})-\sin{(\bold{G}_{i}\bold{r})}\sin{(\bold{g_{i}}\bold{\tau_{b}})} & 0\\
0 & \cos{(\bold{G}_{i}\bold{r})}(1+\cos{(\bold{g_{i}}\bold{\tau_{b}})})+\sin{(\bold{G}_{i}\bold{r})}\sin{(\bold{g_{i}}\bold{\tau_{b}})} \\ 
\end{pmatrix},
\end{align}
\end{widetext}
where $\bold{\tau}_{B}=(0,a\sqrt{3})$ and whose top-layer counterpart has the form,
\begin{align}
\mathcal{E}^{t}=\begin{pmatrix}
\mathcal{E}^{bB} & 0\\
0 & \mathcal{E}^{bA}
\end{pmatrix}.
\end{align}
Upon adding to the tBLG Hamitlonian the relaxation-induced gauge field ($\mathcal{A}_{x}(\vec{r}),\mathcal{A}_{y}(\vec{r})$), the dimerization energy shifts $\mathcal{E}(\vec{r})$ as well as the two strain-induced pseudoscalar pseudopotentials ($V_{z}(\vec{r})=\Upsilon[\partial_{x}\delta U^{h}_{x}(\vec{r})+\partial_{y}\delta U^{h}_{y}(\vec{r})]$ and $V_{||}(\vec{r})=\Upsilon[\partial_{x}\delta U_{x}(\vec{r})+\partial_{y}\delta U_{y}(\vec{r}))$]) the fully relaxed tBLG graphene Hamiltonian written on the four-state basis, $\Psi^{t}_{A},\Psi^{t}_{B},\Psi^{b}_{A},\Psi^{b}_{B}$, becomes
\begin{widetext}
\begin{figure*}[t!]
    \centering
    \includegraphics[width =7in, height=4.5in]{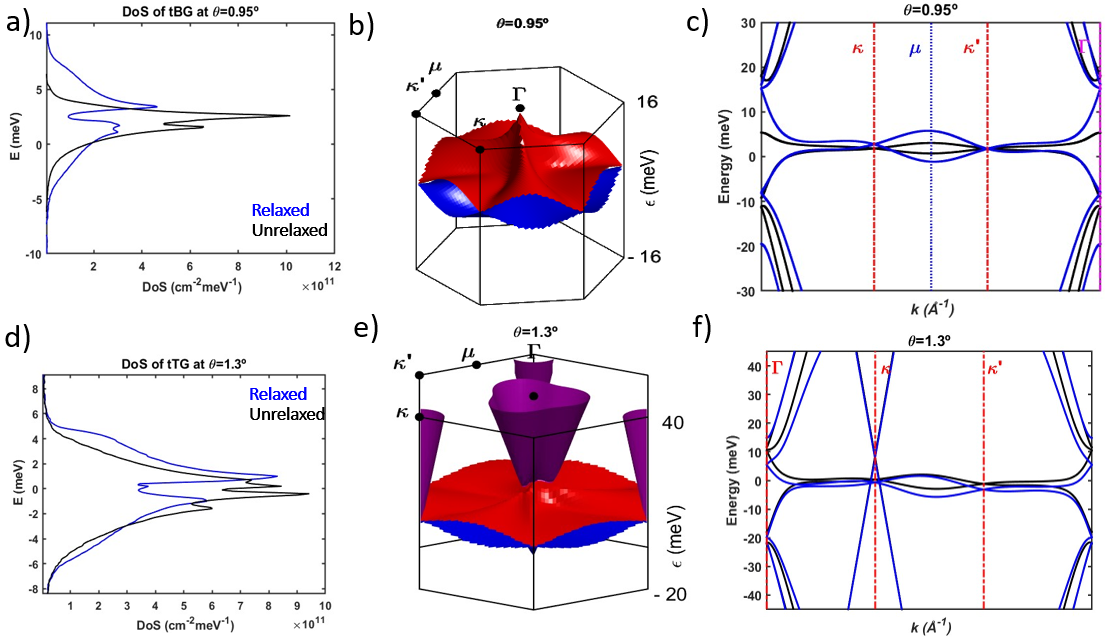}
    \caption{\footnotesize a,c) Comparison of the density of states per valley and the electronic bandstructure of tBG with and without atomic relaxation for an applied twist angle of $\theta=0.95^\circ$. d,f) Comparison of the density of states in one valley and the electronic bandstructure of tTLG with and without atomic relaxation for an applied twist angle of $\theta=1.3^\circ$. Figures b) and e) show the fully relaxed tBLG and tTLG bandstructures. The parameter $\gamma_{2}$ in figure e) was set to zero for clarity purposes.}
\label{fig:bandsstr}
\end{figure*}
\begin{equation}\label{eqn:contBil}
    H_{tBLG}=\begin{pmatrix}
    \Vec{\sigma}_{\theta/2}(-v_{F}i\Vec{\triangledown}+\mathcal{A}(\vec{r}))+\mathcal{E}^{t}(\vec{r})+V_{||}(\vec{r})+V_{z}(\vec{r}) & \mathcal{T}(\vec{r})\\
    \mathcal{T}^{\dagger}(\vec{r}) & \Vec{\sigma}_{-\theta/2}(-v_{F}i\Vec{\triangledown}-\mathcal{A}(\vec{r}))+\mathcal{E}^{b}(\vec{r})-V_{||}(\vec{r})-V_{z}(\vec{r})
    \end{pmatrix}.
\end{equation}
\end{widetext}
Note that in Eq.(\ref{eqn:contBil}), the pseudogauge field $\mathcal{A}(\vec{r})$ and the two pseudoscalar potentials $V_{z}$ and $V_{||}$, have different signs in each layer due to the opposite direction of the atomic displacement field in each graphene membrane. As the twist angle increases, the unit cell becomes smaller suppressing the expansion of the AB and BA regions. This permits us to safely approximate $\vec{u}$ by the expression in Eq.\ref{eqn:Ulin} and $B$ by the formula in Eq.\ref{eqn:Blin}. Due to the large momentum mismatch between nearest-neighboring Dirac cones in the high-twist angle limit, the energy difference between $\bold{K}^{top}$ and $\bold{K}^{bottom}$ becomes larger. As pointed out in Ref.\onlinecite{BernevigPT}, this energetic separation permits to safely truncate the BM Hamiltonian into an $8\times 8$ matrix which only considers the resonant tunneling between a top-layer Dirac cone and its three nearest-neighboring $K$-point states of the bottom layer. Ignoring both the pseudoscalar and the dimerization potential, we account for atomic relaxation by adding a pseudogauge field coupling only the bottom layer $K$-point states (see the inset in Fig.\ref{fig:magic} a)). The resulting Hamiltonian written in the eight-state basis made of 4 Dirac cones ($\Psi^{A_{1}},\Psi^{B_{1}},\Psi^{B_{2}},\Psi^{B_{3}}$) $\times$ 2 sublattice sites ($A,B$) per cone reads as,
\begin{align}\label{eqn:trip}
    H_{8\times 8}=\begin{pmatrix}
   v_{F}\bold{k}\vec{\sigma} & \mathcal{T}^{1} & \mathcal{T}^{2} & \mathcal{T}^{3} \\
   \mathcal{T}^{1\dag} & v_{F}(\bold{k}-\bold{q}_{1})\vec{\sigma} & \mathcal{A}^{1}\vec{\sigma} & \mathcal{A}^{3}\vec{\sigma} \\
    \mathcal{T}^{2\dag} & \mathcal{A}^{2}\vec{\sigma} & v_{F}(\bold{k}-\bold{q}_{2})\vec{\sigma} & \mathcal{A}^{2}\vec{\sigma} \\
    \mathcal{T}^{3\dag} & \mathcal{A}^{3}\vec{\sigma} & \mathcal{A}^{3}\vec{\sigma} & v_{F}(\bold{k}-\bold{q}_{3})\vec{\sigma}
    \end{pmatrix},
\end{align}
where $q_{1}=k_{\theta}(0,1)$, $q_{2}=k_{\theta}(-\sqrt{3}/2,-1/2)$ and $q_{3}=k_{\theta}(\sqrt{3}/2,-1/2)$ are the momentum differences between the $A^{1}$-$B^{1}$,$A^{1}$-$B^{2}$ and $A^{1}$-$B^{3}$ $K-$point states. The terms $\mathcal{A}^{j}$ and $\mathcal{T}^{j}$ in Eq.\ref{eqn:trip} are defined as, 
\begin{align}\label{eqn:matrixdef}
    &
    \mathcal{T}^{j}=\omega_{0}\sigma_{0}+\omega_{1}\Big[\cos{\Big(\frac{2\pi}{3}(j-1)\Big)}\sigma_{x}+\sin{\Big(\frac{2\pi}{3}(j-1)\Big)}\sigma_{y}\Big], \\ \nonumber
    &
    \mathcal{A}^{j}\vec{\sigma}=v_{F}\begin{pmatrix}
        0 & Ae^{-i\pi/2}e^{-ij2\pi/3} \\
        Ae^{i\pi/2}e^{ij2\pi/3} & 0 \\
    \end{pmatrix}, \nonumber
\end{align}
The $C_{3}$ invariance of the Hamiltonian in Eq.\ref{eqn:trip} (commonly termed ``the tripod Hamiltonian") imposes that
\begin{align}
\hat{C}^{\dag}_{3}\hat{H}(R_{2\pi/3}[\vec{r}])\hat{C}_{3}=\hat{H}[\bold{A}],
\end{align}
where $R_{2\pi/3}$ is the rotation operator by 120º and $\hat{C}_{3}\equiv e^{\frac{i2\pi}{3}\sigma_{z}}$. Using the perturbative approach presented in Appendix. \ref{app:twoband}, we formulated an effective low-energy model of tBLG\cite{bennett2023twisted} projecting the high-energy bands in Eq.\ref{eqn:trip} into the lowest-energy states. The commonly used formula for the renormalized Fermi velocity $v^{*}=v_{F}\frac{(1-3\alpha^{2})}{(1+6\alpha^{2})}, \alpha=\omega/(v_{F}\Delta K)$ for the isotropic model of tBLG\cite{BM11} ($\omega_{1}=\omega_{0}$) becomes  
\begin{align}\label{eqn:eqfermibil}
    v^{*}\approx v_{F}\frac{\Big(1-\frac{3\omega_{1}^{2}}{(v_{F}A+v_{F}\Delta K)^{2}}\Big)}{\Big(1+\frac{6\omega_{1}^{2}}{(v_{F}A+v_{F}\Delta K)^{2}}\Big)},
\end{align}
revealing that the effect of atomic relaxation is no other than an effective increase in $\Delta K$. This implies that atomic reconstruction effectively increases the twist angle an amount $\delta \theta=A/K$. Within the linear response theory framework previously presented, the gauge field magnitude is $A=\Big(-\frac{3}{4}\frac{t|\beta |}{v_F}\frac{V_{vdW}}{\mu\theta}\Big)$ which means an effective increase in twist angle by $\delta \theta=\Big(\frac{9\pi}{16}\frac{t|\beta |}{v_F}\frac{V_{vdW}a}{\mu\theta}\Big)$ as clearly shown in see Fig.\ref{fig:magic} a). \\
\indent The continuum formalism previously developed for twisted bilayer graphene can be generalized to the twisted trilayer structure presented in Fig.\ref{fig:bandsstr} b). In tTLG, the top and bottom layer graphene dispersions are described by two Dirac Hamiltonians rotated by an angle $-\theta/2$. Conversely, the middle layer term consists of a Dirac cone rotated an amount $\theta/2$ in the opposite direction to its top and bottom layer counterpart. While the interlayer tunneling terms connecting neighboring layers remain the same as in tBLG, atomic relaxation leads to significant differences compared to tBLG. Firstly, due to the $z\longleftrightarrow-z$ symmetry of the twisted trilayer structure, the twisted middle layer does not experience any out-of-plane atomic relaxation while the outer ones do. Therefore, the scalar potential $V_{z}(\vec{r})$ originated from the out-of-plane superlattice corrugation is only included at the very top and bottom layer. However, given the enhanced in-plane atomic distortion of the middle layer, the scalar potential induced by the adhesion energies $V_{||}(\vec{r})$ as well as the pseudogauge field $\bold{\mathcal{A}}(\vec{r})$, is twice as large in the middle layer as compared to the top and bottom ones. Finally, we account for the energy difference between the dimer and non-dimer pairs formed between the top and bottom graphene layers. Given that both the top and bottom layers lie perfectly AA-aligned to each other, we introduce into the tTLG Hamiltonian a hybridization term connecting the overlaying carbon atoms. This term usually labeled as $\gamma_{2}/2=8.5$meV[\onlinecite{yingraphite}] in the Slonczewski-Weiss-McClure terminology, connects the very top and bottom layer via the AA and BB top and bottom layer sublattices,
\begin{align}
    \mathcal{G}=\begin{pmatrix}
    \gamma_{2}/2 & 0\\
    0 & \gamma_{2}/2 
    \end{pmatrix}.
\end{align}
As noted in Ref.\onlinecite{gam2Young,game2Youngrombohedral}, a negative $\gamma_{2}$ fits better the electronic bandstructure of an ABA trilayer while a positive $\gamma_{2}$ describes more accurately the electronic structure of an ABC (rombohedral) trilayer graphene. Since our trilayer system hybridizes two AA-stacked graphene monolayers, a negative sign will be used for our calculations. The full tTLG Hamiltonian can therefore be written in a six-state basis $\Psi^{t}_{A},\Psi^{t}_{B},\Psi^{m}_{A},\Psi^{m}_{B},\Psi^{b}_{A},\Psi^{b}_{B}$ as,
\begin{widetext}
\begin{align}\footnotesize\label{eqn:trilcont}
    H_{tTLG}=\begin{pmatrix}
    \Vec{\sigma}_{-\frac{\theta}{2}}(-v_{F}i\Vec{\bold{\triangledown}}+\mathcal{A}(\vec{r}))+\mathcal{E}
    (\vec{r})+V_{t}(\vec{r}) & \mathcal{T}(\vec{r}) & \mathcal{G} \\
    \mathcal{T}^{\dagger}(\vec{r}) & \Vec{\sigma}_{\frac{\theta}{2}}(-v_{F}i\Vec{\triangledown}-2\mathcal{A}(\vec{r}))+2\mathcal{E}^{\dagger}(\vec{r})-2V_{t}(\vec{\vec{r}}) & \mathcal{T}^{\dagger}(\vec{r}) \\
    \mathcal{G} & \mathcal{T}(\vec{r}) & \Vec{\sigma}_{-\frac{\theta}{2}}(-v_{F}i\Vec{\triangledown}+\mathcal{A}(\vec{r}))+\mathcal{E}(\vec{r})+V_{t}(\vec{\vec{r}})
    \end{pmatrix}.
\end{align}
\end{widetext}
\indent Atomic relaxation in tBLG was found to significantly alter its band structure. Firstly, the pseudoscalar potentials were responsible for a slight energy shift between the $\kappa$ and $\kappa'$ points while the pseudomagnetic field widened the $\Gamma$-point bandwidth (see Fig.\ref{fig:magic} c)). Furthermore, the dimer-non dimer term $\Delta$ shifted the middle bands towards higher energy leading to a greater electron-hole asymmetry appreciable in the density of states (see Fig.\ref{fig:bandsstr} d)). \\ 
\indent Due to an increase in the local twist angle in the AA regions, we found that a smaller twist angle of the order of $0.8-0.9^\circ$ was necessary to flatten the middle band of tBLG. Moreover, between $0.8-0.9^\circ$ nearly constant pseudomagnetic fields of approximately 30T were found very localized in the vicinity of the AA-regions. These pseudomagnetic fields heavily suppress the Dirac dispersion since the electron wavefunctions of the middle band are also confined around the AA patches where the pseudomagnetic field maxima are located. This additional confining force leads to a very stable flat-band plateau for twist angles ranging between $0.8-0.85^\circ$. Likewise, the magic angle of fully relaxed tTLG also shifted towards lower angles and formed a magic-angle plateau between $1-1.1^\circ$ due to the same confining mechanism as in tBLG. Nonetheless, due to the enhanced pseudomagnetic fields in the middle layer AA regions, the flat-band plateau becomes even more pronounced than in tBLG.

\section{Conclusions}

We have analyzed in detail the terms that modify the standard model used to study twisted graphene bilayers and symmetrically twisted graphene trilayers. 

$\bullet$ We present an analytic scheme, based on linear response, to study the in-plane lattice relaxation, based on two parameters, the Lamé elastic coefficient $\mu$, and the difference in the adhesion potential between the $AA$ and $AB$ regions of a graphene bilayer, $V_{VdW}$. These two parameters lead to a dimensionless number $V_{vdW}/ \mu \sim 10^{-4}$ which quantifies the angle dependence of the in-plane deformations and the strength of the resulting pseudomagnetic fields acting on the electrons. It is worth noting that this dimensionless parameter also characterized the competition between the elastic and the adhesion forces in other situations such as in the shape of bubbles\cite{Ketal16} or in mechanically deformed nanoribbons\cite{psg23}. Within the linear response approximation, the magnitude of the pseudomagnetic fields is roughly independent of the twist angle $B \approx 40$T. For angles similar or lower to the first magic angle, $\theta \lesssim 1^\circ$, we computed the in-plane atomic displacements by going beyond linear response theory. We have also formulated an effective two-band Hamiltonian capable of quantifying the shift in magic angle due to atomic relaxation. \\
\indent $\bullet$ The in-plane atomic displacements in the central layer of a symmetric twisted trilayer graphene\cite{KKTV19} is increased by a factor of $2$ with respect to the relaxation in a twisted bilayer, as the central layer experiences forces of the same sign from the two surface layers. In contrast to tBLG, atomic relaxation in twisted trilayer graphene significantly stabilizes the flat-band plateau around $1^\circ$ while having two very energetically separated regimes, a middle flat-band similar to the one in tBLG and a highly dispersive Dirac cone above charge neutrality. \\
\indent $\bullet$ We formulated a simple model based on a few harmonics to describe the pseudoscalar potential due to the change in the interlayer distance. Together with the pseudoscalar field produced by the adhesion forces, the energetic degeneracy between the $\kappa$ and $\kappa'$ points in the Moiré Brillouin zone was lifted by a few meV. \\
\indent $\bullet$ Finally, we analyzed the effect of a sublattice-dependent potential\cite{FullSWC} $\Delta\sim 25$meV which is known to play a role in the band structure of Bernal bilayers. We found that this term energetically shifted the flat middle band towards higher energies, producing a visibly particle-hole asymmetric density of states profile. 

\section{Acknowledgments}

We thank José Ángel Silva-Guillén, Zhen Zhan and Pierre A. Pantaleón for fruitful discussions. F.Guinea and A.Ceferino acknowledge the support from the Graphene Flagship, Core 3, grant number 881603 and from grants NMAT2D (Comunidad de Madrid, Spain), SprQuMat and (MAD2D-CM)-MRR
MATERIALES AVANZADOS-IMDEA-NC. We also acknowledge support from NOVMOMAT, Grant PID2022-142162NB-I00 funded by
MCIN/AEI/ 10.13039/501100011033 and, by “ERDF A way of making Europe”. IMDEA Nanociencia acknowledges support from the “Severo Ochoa” Programme for Centres of Excellence in R\&D
(Grant No. SEV-2016-0686).

\bibliography{bibli}

\appendix

\begin{widetext}
    
\section{Elastic calculations in twisted bilayer graphene}\label{app:elasticity}

To obtain the relaxation-induced atomic displacement field in tBLG, we start with an energy functional describing the sum of the elastic and the adhesion energy,
\begin{align}
&
U=E_{el}+E_{ad}=\sum_{n=t,b}\int \Bigg(\Bigg[\frac{\kappa}{2}\Big(\frac{\partial h^{(n)}}{\partial x^{2}}+\frac{\partial  h^{(n)} }{\partial y^{2}}\Big)^{2}+\frac{(\mu+\lambda)}{2}\Big( u^{(n)}_{xx}+ u^{(n)}_{yy}\Big)^{2}+\frac{\mu}{2}\Bigg(\Big(u^{(n)}_{xx}-u^{(n)}_{yy}\Big)^{2}+\Big(u^{(n)}_{yx}+u^{(n)}_{xy}\Big)^{2}\Bigg)\Bigg] \\ \nonumber
&
+\Bigg[\sum^{i=3}_{i=1}V_{vdW}\cos{(\bold{G}_{i}\bold{r}+\bold{g}_{i}\cdot(\bold{u}^{1}-\bold{u}^{2}))}\Bigg]\Bigg) \bold{dR}. \\ 
\nonumber
\end{align}
where $\bold{G}=-\theta\hat{z}\times\bold{g}$.Due to a net zero force in each point of the two membranes, we require $u^{t}=-u^{b}$ and $h^{t}=-h^{b}$. Therefore, we obtain the following equation for the total energy as a function of $u\equiv u^{t}$ 
\begin{align}
&
U=2E_{el}+E_{ad}=\int \Bigg(2\Bigg[\frac{\kappa}{2}\Big(\frac{\partial h^{(1)}}{\partial x^{2}}+\frac{\partial  h^{(1)} }{\partial y^{2}}\Big)^{2}+\frac{(\mu+\lambda)}{2}\Big( u^{(1)}_{xx}+ u^{(1)}_{yy}\Big)^{2}+\frac{\mu}{2}\Bigg(\Big(u^{(1)}_{xx}-u^{(1)}_{yy}\Big)^{2}+\Big(u^{(1)}_{yx}+u^{(1)}_{xy}\Big)^{2}\Bigg)\Bigg] \\ \nonumber
&
+\Bigg[\sum^{i=3}_{i=1}V_{vdW}\cos{(\bold{G}_{i}\bold{r}+2\bold{g}_{i}\cdot\bold{u})}\Bigg]\Bigg) \bold{dR}, \\ 
\nonumber
\end{align}
where
\begin{align}
    u_{\alpha\beta}=\frac{1}{2}\Big(\partial_{\beta}u_{\alpha}+\partial_{\alpha}u_{\beta}+\partial_{\alpha}h\partial_{\beta}h\Big),
\end{align}
and
\begin{align}
h(\Vec{\bold{r}})=\Big(d+\frac{3h_{0}}{2}\Big)+h_{0}\sum^{i=3}_{i=1}\cos{\Big(\Vec{G}_{i}\bold{\cdot}\Vec{r}\Big)}.
\end{align}
To minimize the energy, we use the generalized Euler-Lagrange equations for the field $\vec{u}=(u_{x},u_{y})$,
\begin{align}
    &
    \frac{\partial L}{\partial u_{x}}-\frac{\partial}{\partial x}\frac{\partial L}{\partial (\partial_{x}u_{x})}-\frac{\partial}{\partial y}\frac{\partial L}{\partial (\partial_{y}u_{x})}=0, \\ \nonumber
    &
    \frac{\partial L}{\partial u_{y}}-\frac{\partial}{\partial x}\frac{\partial L}{\partial (\partial_{x}u_{y})}-\frac{\partial}{\partial y}\frac{\partial L}{\partial (\partial_{y}u_{y})}=0, \nonumber
\end{align}
resulting in the following two equations,
\begin{align}\label{eqn:prev}
    &
    (\mu+\lambda)\frac{\partial_{x}}{2}\Big(2\partial_{x}u_{x}+(\partial_{x}h)^{2}+2\partial_{y}u_{y}+(\partial_{y}h)^{2}\Big)+\mu\frac{\partial_{x}}{2}\Big(2\partial_{x}u_{x}+(\partial_{x}h)^{2}-2\partial_{y}u_{y}-(\partial_{y}h)^{2}\Big) \\ \nonumber
    &
    +\mu\partial_{y}\Big(\partial_{x}u_{y}+\partial_{y}u_{x}+\partial_{y}h\partial_{x}h\Big)=-\sum^{i=3}_{i=1}V_{vdW}\sin{(\Vec{G}_{i}\Vec{r}+2\Vec{g}_{i}\Vec{u})}g_{ix},\\ \nonumber
    &
    (\mu+\lambda)\frac{\partial_{y}}{2}\Big(2\partial_{x}u_{x}+(\partial_{x}h)^{2}+2\partial_{y}u_{y}+(\partial_{y}h)^{2}\Big)+\mu\frac{\partial_{y}}{2}\Big(-2\partial_{x}u_{x}-(\partial_{x}h)^{2}+2\partial_{y}u_{y}+(\partial_{y}h)^{2}\Big) \\ \nonumber
    &
    +\mu\partial_{x}\Big(\partial_{x}u_{y}+\partial_{y}u_{x}+\partial_{y}h\partial_{x}h\Big)=-\sum^{i=3}_{i=1}V_{vdW}\sin{(\Vec{G}_{i}\Vec{r}+2\Vec{g}_{i}\Vec{u})}g_{iy}.\nonumber
\end{align}
\begin{figure}[t!]
    \centering
    \includegraphics[width =7.5in, height=3.9in]{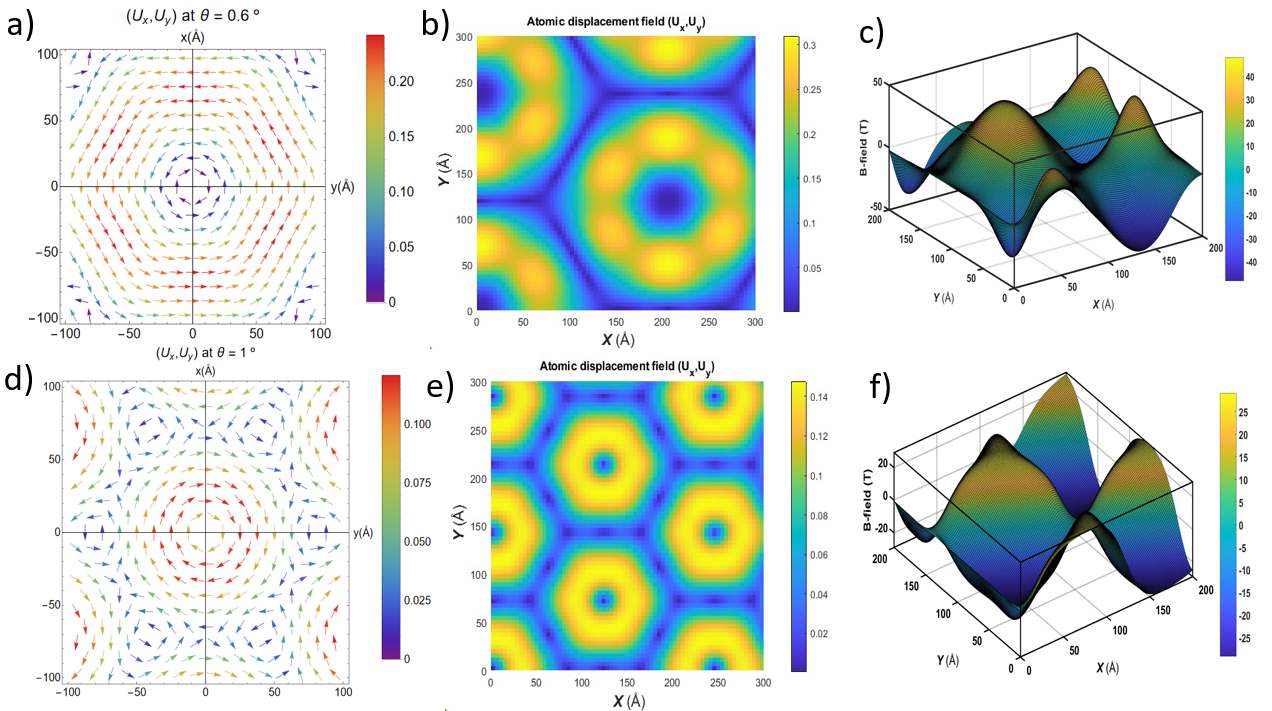}
    \caption{\footnotesize  a) Strain field of the top graphene layer calculated using Eq.(\ref{eqn:NonlinU}) and (\ref{eqn:NonlinUy}) for an applied twist angle of $\theta=0.6^\circ$ in units of \AA. b-c) Strain field magnitude in units of \AA and corresponding pseudomagnetic field at $\theta=0.6^\circ$ in units of Teslas calculated using the self-consistent method proposed by Nam and Koshino in Ref.\onlinecite{NK17} with the parameters shown in Table.\ref{tbl:param}. d) Strain field of the top graphene layer calculated using Eq.(\ref{eqn:NonlinU}) and (\ref{eqn:NonlinUy}) for an applied twist angle of $\theta=1^\circ$.e-f) Strain field and corresponding pseudomagnetic field at $\theta=1^\circ$ calculated using the self-consistent method in Ref.\onlinecite{NK17}.}
\label{fig:comparison}
\end{figure}
By simply Fourier transforming Eq.(\ref{eqn:prev}), we obtain
\begin{align}\label{eqn:Fourier}
    &
    \begin{pmatrix}
    \delta \bold{U}_{x}(\Vec{q}) \\
    \delta \bold{U}_{y}(\Vec{q}) \\
    \end{pmatrix}=\frac{\begin{pmatrix}
    (2\mu+\lambda)q^{2}_{y}+\mu q^{2}_{x} & -(\mu+\lambda) q_{x}q_{y}\\
     -(\mu+\lambda)q_{x}q_{y} & (2\mu+\lambda)q^{2}_{x}+\mu q^{2}_{y}
    \end{pmatrix}}{[(2\mu+\lambda)q^{2}_{x}+\mu q^{2}_{y}][(2\mu+\lambda)q^{2}_{y}+\mu q^{2}_{x}]-((\mu+\lambda)(q_{x}q_{y}))^{2}}\\ \nonumber
    &
    \sum_{i,j}\begin{pmatrix} \\ \nonumber
     &
     V_{vdW}g_{ix}\frac{(\delta(\Vec{q}-\Vec{G}_{i})-\delta(\Vec{q}+\Vec{G}_{i}))}{2i}+[(2\mu+\lambda)h^{2}_{0}G^{2}_{ix}G_{jx}+(\mu+\lambda) h^{2}_{0}G_{iy}G_{ix}G_{jy}+h^{2}_{0}\mu G^{2}_{iy}G_{jx}]\\ \nonumber
     &
     \times \frac{\Big(\delta(\Vec{q}-\Vec{G}_{i}-\Vec{G}_{j})-\delta(\Vec{q}-\Vec{G}_{i}+\Vec{G}_{j})+\delta(\Vec{q}+\Vec{G}_{i}-\Vec{G}_{j})-\delta(\Vec{q}+\Vec{G}_{i}+\Vec{G}_{j})\Big)}{4i} \\ \nonumber
     &
     \\ \nonumber
     &
     V_{vdW}g_{iy}\frac{(\delta(\Vec{q}-\Vec{G}_{i})-\delta(\Vec{q}+\Vec{G}_{i}))}{2i}+[(2\mu+\lambda)h^{2}_{0}G^{2}_{iy}G_{jy}+(\mu+\lambda) h^{2}_{0}G_{ix}G_{iy}G_{jx}+h^{2}_{0}\mu G^{2}_{ix}G_{jy}]\\ \nonumber
     &
     \times \frac{\Big(\delta(\Vec{q}-\Vec{G}_{i}-\Vec{G}_{j})-\delta(\Vec{q}-\Vec{G}_{i}+\Vec{G}_{j})+\delta(\Vec{q}+\Vec{G}_{i}-\Vec{G}_{j})-\delta(\Vec{q}+\Vec{G}_{i}+\Vec{G}_{j})\Big)}{4i} \\ \nonumber
    \end{pmatrix},
    \end{align}
which, upon inverse Fourier transform becomes
\begin{align}\label{eqn:linearalg}
    &
    \delta U_{x}(\bold{x})=\sum_{i=1-3}\frac{\Big(V_{vdW}\Big((2\mu+\lambda)G^{2}_{iy}+\mu G^{2}_{ix}\Big)g_{ix}+\Big(-V_{vdW}(\mu+\lambda) G_{ix}G_{iy}\Big)g_{iy}\Big)\sin{(\Vec{G}_{i}\Vec{r})}}{\mu(2\mu+\lambda)G^{4}}+\sum_{i=1-6,j=1-3}\sin{[(\Vec{G}_{i}+\Vec{G}_{j})\Vec{r}]}\times\\ \nonumber
    &
    \Big[\frac{1}{2}\frac{\Big([(2\mu+\lambda)h^{2}_{0}G^{2}_{ix}G_{jx}+(\mu+\lambda)h^{2}_{0}G_{iy}G_{ix}G_{jy}+h^{2}_{0}\mu G^{2}_{iy}G_{jx}]\Big)((2\mu+\lambda)(G_{iy}+G_{jy})^{2}+\mu (G_{ix}+G_{jx})^{2})}{\mu(2\mu+\lambda)((G_{ix}+G_{jx})^{2}+(G_{iy}+G_{jy})^{2})^{2}} \\ \nonumber
    &
    -\frac{1}{2}\frac{\Big([(2\mu+\lambda)h^{2}_{0}G^{2}_{iy}G_{jy}+(\mu+\lambda)h^{2}_{0}G_{ix}G_{iy}G_{jx}+h^{2}_{0}\mu G^{2}_{ix}G_{jy}]\Big)(\mu+\lambda)(G_{iy}+G_{jy})(G_{ix}+G_{jx})}{\mu(2\mu+\lambda)((G_{ix}+G_{jx})^{2}+(G_{iy}+G_{jy})^{2})^{2}}\Big], \nonumber
\end{align}
with $\delta U_{y}(\bold{x})$ defined as,
\begin{align}\label{eqn:linearaly}
    \delta U_{y}(\bold{x})=\delta U_{x}(\bold{x})[G_{ix}\longrightarrow G_{iy},G_{iy}\longrightarrow G_{ix}].
\end{align}
Since the $h^{2}_{0}$ term is much smaller than the $V_{vdW}$ term, we can simplify Eq.(\ref{eqn:linearalg}) as
 \begin{align} \label{eqn:simple}
    \vec{u}\approx\frac{V_{vdW}}{\mu G^2}   \sum_{i=1,2,3} \vec{g}_i \sin ( \vec{G}_i \vec{r} ).     
 \end{align}
While Eq.(\ref{eqn:linearalg}),(\ref{eqn:linearaly}) correctly captures the membrane deformations at very large angles, we can improve this model by plugging Eq.(\ref{eqn:simple}) into the adhesion term and apply the Jacobi-Anger identity,
 \begin{align} \label{eqn:Jacobi}
&
\cos{(z\sin{(\theta )})}=J_0(z)+2\sum _{n=1}^{\infty } J_{2n}(z) \cos (\text{2n$\theta $}), \\ \nonumber
&
\sin{(z\sin(\theta ))}=2 \sum _{n=1}^{\infty } J_{2n-1}(z) \sin (\text{(2n-1)$\theta $}). \nonumber
 \end{align}
The adhesion potential then becomes,
\begin{align}
 &
 \sum^{i=3}_{i=1}V_{vdW}\sin{(\Vec{G}_{i}\Vec{r}+2\Vec{g}_{i}\Vec{u})}=V_{vdW}\sum^{i=3}_{i=1}  \sin{(\vec{G}_{i}\vec{r})}\Bigg[\sum^{\infty}_{l=0}\Big(J^{1}_{l}\cos{(2l\vec{G}_{1}\vec{r})}\Big)\sum^{\infty}_{m=0}\Big(J^{2}_{m}\cos{(2m\vec{G}_{2}\vec{r})}\Big)\sum^{\infty}_{n=0}\Big(J^{3}_{n}\cos{(2n\vec{G}_{3}\vec{r})}\Big) \\ \nonumber
 &
 -\Big(\sum^{\infty}_{l=0}J^{1}_{l}\cos{(2l\vec{G}_{1}\vec{r})}\Big)\Big(\sum^{\infty}_{m=1} J^{2}_{2m-1}\sin{((2m-1)\vec{G}_{2}\vec{r})}\Big)\Big(\sum^{\infty}_{n=1} J^{3}_{2n-1}\sin{((2n-1)\vec{G}_{3}\vec{r})}\Big) \\ \nonumber
 &
 -\Big(\sum^{\infty}_{l=1} J^{1}_{2l-1}\sin{((2l-1)\vec{G}_{1}\vec{r})}\Big)\Big(\sum^{\infty}_{m=0} J^{2}_{m}\cos{(2m\vec{G}_{2}\vec{r})}\Big)\Big(\sum^{\infty}_{n=1} J^{3}_{2n-1}\sin{((2n-1)\vec{G}_{3}\vec{r})}\Big) \\ \nonumber
 &
 -\Big(\sum^{\infty}_{l=1} J^{1}_{2l-1}\sin{((2l-1)\vec{G}_{1}\vec{r})}\Big)\Big(\sum^{\infty}_{m=1} J^{2}_{2m-1}\sin{((2m-1)\vec{G}_{2}\vec{r})}\Big)\Big(\sum^{\infty}_{n=0} J^{3}_{n}\cos{(2n\vec{G}_{3}\vec{r})}\Big)\Bigg]+ \\ \nonumber
 &
 \sum^{i=3}_{i=1}V_{vdW}\cos{(\vec{G}_{i}\vec{r})}\Bigg[\Big(\sum^{\infty}_{l=1} J^{1}_{2l-1}\sin{((2l-1)\vec{G}_{1}\vec{r})}\Big)\Big(\sum^{\infty}_{m=0} J^{2}_{m}\cos{(2m\vec{G}_{2}\vec{r})}\Big)\Big(\sum^{\infty}_{n=0}J^{3}_{n}\cos{(2n\vec{G}_{3}\vec{r})}\Big) \\ \nonumber
 &
 +\Big(\sum^{\infty}_{l=0} J^{1}_{l}\cos{(2l\vec{G}_{1}\vec{r})}\Big)\Big(\sum^{\infty}_{m=1} J^{2}_{2m-1}\sin{((2m-1)\vec{G}_{2}\vec{r})}\Big)\Big(\sum^{\infty}_{n=0}J^{3}_{n}\cos{(2n\vec{G}_{3}\vec{r})}\Big) \\ \nonumber
 &
 +\Big(\sum^{\infty}_{l=0}J^{1}_{l}\cos{(2l\vec{G}_{1}\vec{r})}\Big)\Big(\sum^{\infty}_{m=0}J^{2}_{m}\cos{(2m\vec{G}_{2}\vec{r})}\Big)\Big(\sum^{\infty}_{n=1}J^{3}_{2n-1}\sin{((2n-1)\vec{G}_{3}\vec{r})}\Big) \\ \nonumber
 &
 -\Big(\sum^{\infty}_{l=1} J^{1}_{2l-1}\sin{((2l-1)\vec{G}_{1}\vec{r})}\Big)\Big(\sum^{\infty}_{m=1} J^{2}_{2m-1}\sin{((2m-1)\vec{G}_{2}\vec{r})}\Big) \Big(\sum^{\infty}_{n=1} J^{3}_{2n-1}\sin{((2n-1)\vec{G}_{3}\vec{r})}\Big) \Bigg],
 \end{align}
where $J^{1}_{l},J^{2}_{m},J^{3}_{n}$, are the Bessel prefactors in Eq.\ref{eqn:Jacobi} with prefactors $\Big(\frac{2V_{vdW}}{\mu G^{2}}\vec{g}_{1}\cdot \vec{g}_{i}\Big),\Big(\frac{2V_{vdW}}{\mu G^{2}}\vec{g}_{2}\cdot \vec{g}_{i}\Big)$ and $\Big(\frac{2V_{vdW}}{\mu G^{2}}\vec{g}_{3}\cdot \vec{g}_{i}\Big)$ respectively. The above expression can be simplified as,
 \begin{align}
 &
 \sum^{i=3}_{i=1}V_{vdW}\sin{(\Vec{G}_{i}\Vec{r}+2\Vec{g}_{i}\Vec{u})} =\sum^{3,\infty\infty,\infty}_{i=1,l=0,m=0,n=0}A^{\pm l,\pm m,\pm n,i}\sin{(\vec{G}_{i}\vec{r}\pm l\vec{G}_{1}\vec{r}\pm m\vec{G}_{2}\vec{r}\pm n\vec{G}_{3}\vec{r})},
 \end{align}
 where
 \begin{align}\label{eqn:Adeff}
     &
    \vert A^{\pm l,\pm m,\pm n,i} \vert=V_{vdW}J_{l}\Big(\frac{2V_{vdW}}{\mu G^{2}}\vec{g}_{1}\cdot \vec{g}_{i}\Big)J_{m}\Big(\frac{2V_{vdW}}{\mu G^{2}}\vec{g}_{2}\cdot \vec{g}_{i}\Big)J_{n}\Big(\frac{2V_{vdW}}{\mu G^{2}}\vec{g}_{3}\cdot \vec{g}_{i}\Big), \\ \nonumber
    &
    A^{\pm l,\pm m,\pm n,i}=\vert A^{l,m,n,i}\vert Sign(\pm l,\pm m,\pm n), \\ \nonumber
 \end{align}
 and $Sign(l,m,n)$ is defined as,
\begin{align}\label{eqn:Adefth}
    &
    \bullet Sign(l,m,n)= 1 \textit{ for l,m,n even}. \\ \nonumber
    &
    \bullet Sign(l,m,n)=(-1)(-1)sgn(p)sgn(q) \textit{ for 2 odd indices p and q }.\\ \nonumber
    &
    \bullet Sign(l,m,n)= sgn(t) \textit{ for 1 odd indices, t being the odd index  }. \\ \nonumber
    &
    \bullet Sign(l,m,n)= (-1)(-1)sgn(l)\times sgn(m)\times sgn(n)\textit{ for all odd indices}. \nonumber
 \end{align} 
  In defining $A_{x},A_{y}$ as,
 \begin{align}\label{eqn:Adefs}
    &
    A^{l,m,n,i}_{x}=A^{\pm l,\pm m,\pm n,i}g_{x}, \\ \nonumber
    &
    A^{l,m,n,i}_{y}=A^{\pm l,\pm m,\pm n,i}g_{y},
 \end{align}
 we can write the full solution for $(\delta U_{x},\delta U_{y})$ as,
 \begin{align}
    &
    \begin{pmatrix}
    \delta \bold{U}_{x}(\Vec{q}) \\
    \delta \bold{U}_{y}(\Vec{q}) \\
    \end{pmatrix}=\frac{\begin{pmatrix}
    (2\mu+\lambda)q^{2}_{y}+\mu q^{2}_{x} & -(\mu+\lambda) q_{x}q_{y}\\
     -(\mu+\lambda)q_{x}q_{y} & (2\mu+\lambda)q^{2}_{x}+\mu q^{2}_{y}
    \end{pmatrix}\times}{[(2\mu+\lambda)q^{2}_{x}+\mu q^{2}_{y}][(2\mu+\lambda)q^{2}_{y}+\mu q^{2}_{x}]-((\mu+\lambda)(q_{x}q_{y}))^{2}}\\ \nonumber
    &
    \begin{pmatrix}
     &
    \sum^{3,\infty\infty,\infty}_{i=1,l=0,m=0,n=0}A^{l,m,n,i}_{x}\frac{(\delta(\Vec{q}-(\pm l\vec{G}_{1}\pm m\vec{G}_{2}\pm n\vec{G}_{3}+\vec{G}_{i}))-\delta(\Vec{q}+(\pm l\vec{G}_{1}\pm m\vec{G}_{2}\pm n\vec{G}_{3}+\vec{G}_{i}))}{2i} \\ \nonumber
     &
     \\ \nonumber
     &
    \sum^{3,\infty\infty,\infty}_{i=1,l=0,m=0,n=0}A^{l',m',n',i'}_{y}\frac{(\delta(\Vec{q}-(\pm l'\vec{G}_{1}\pm m'\vec{G}_{2}\pm n'\vec{G}_{3}+\vec{G}_{i'}))-\delta(\Vec{q}+(\pm l'\vec{G}_{1}\pm m'\vec{G}_{2}\pm n'\vec{G}_{3}+\vec{G}_{i'})))}{2i}\\ \nonumber
    \end{pmatrix},
    \end{align}
leading to
\begin{align}
    &
    \delta U_{x}(\vec{x})=\frac{\sum^{3,\infty\infty,\infty}_{i=1,l=0,m=0,n=0}((2\mu+\lambda)G_{l,m,n,iy}^{2}+\mu G_{l,m,n,ix}^{2})A^{\pm l,\pm m,\pm n,i}_{x}\sin{((\pm l\vec{G}_{1}\pm m\vec{G}_{2}\pm n\vec{G}_{3}+\vec{G}_{i})\vec{r})}}{\mu(2\mu+\lambda)G_{l,m,n,i}^{4}}\\ \nonumber
    &
    -\sum^{3,\infty\infty,\infty}_{i'=1,l'=0,m'=0,n'=0}\frac{(\mu+\lambda)G_{l',m',n',i'y}G_{l',m',n',i'x}A^{\pm l',\pm m',\pm n',i'}_{y}\sin{(\pm l'\vec{G}_{1}\vec{r}\pm m'\vec{G}_{2}\vec{r}\pm n'\vec{G}_{3}\vec{r}+\vec{G}_{i'}\vec{r})}}{\mu(2\mu+\lambda)G_{l',m',n',i'}^{4}}, \\ \nonumber
    &
    \delta U_{y}(\bold{x})=\delta U_{x}(\bold{x})[G_{ix}\longrightarrow G_{iy},G_{iy}\longrightarrow G_{ix}]. \nonumber
\end{align}

\section{Effective continuum model of fully relaxed \MakeLowercase{t}BLG}\label{app:continuumapend}

In this Appendix, we review how the gauge field $\mathcal{A}$ is derived from the strain field $(\delta U_{x},\delta U_{y})$ following the analysis presented in Ref.\onlinecite{xie2023lattice} and \onlinecite{VKG10} and we discuss the effect of atomic relaxation in the interlayer hoppings.
We start from the usual Slater-Koster form of the nearest-neighboring $p_{z}$ orbital hopping,
\begin{align}
    t(d)=V_{\sigma}\Big[\frac{\bold{d}\cdot\bold{\hat{z}}}{d}\Big]+V_{\pi}\Big[1-\Big(\frac{\bold{d}\cdot\bold{\hat{z}}}{d}\Big)^{2}\Big],  
\end{align}
where $V_{\sigma}(R)=V^{0}_{\sigma}e^{-(\frac{|R|-a_{cc}}{r_{0}})}$, $V_{\pi}(R)=V^{0}_{\pi}e^{-(\frac{|R|-d}{r_{0}})}$, $d$ is the interlayer distance and $a_{cc}=a/\sqrt{3}$ is the distance between two neighboring carbon atoms in a graphene membrane. Using the monolayer graphene lattice vectors $\vec{a}_{1}=a\Big(\frac{1}{2},\frac{\sqrt{3}}{2}\Big),\vec{a}_{2}=a\Big(-\frac{1}{2},\frac{\sqrt{3}}{2}\Big)$ and following the analysis presented in Ref.\onlinecite{xie2023lattice}, we evaluate down to first order in $u$ the change in the three nearest-neighboring intralayer $p_{z}$ orbital hopping,
\begin{align}
    &
    \delta t_{1}=-t\beta u_{yy}+O(u^{2}), \\ \nonumber
    &
    \delta t_{2}=-t\beta \Big(\frac{3}{4}u_{xx}+\frac{\sqrt{3}}{4}u_{xy}+\frac{\sqrt{3}}{4}u_{yx}+\frac{1}{4}u_{yy}\Big)+O(u^{2}), \\ \nonumber
    &
    \delta t_{3}=-t\beta\Big(\frac{3}{4}u_{xx}-\frac{\sqrt{3}}{4}u_{xy}-\frac{\sqrt{3}}{4}u_{yx}+\frac{1}{4}u_{yy}\Big)+O(u^{2}), \\ \nonumber
\end{align}
where $\beta=\frac{a_{cc}}{r_{0}}$ and $t=-\gamma_{0}=-2.6$eV. The tight-binding Hamiltonian of monolayer graphene is well known to read as
\begin{align}
    H_{G}=\begin{pmatrix}
    0 & g(k)\\
    g(k) & 0
    \end{pmatrix},
\end{align}
where $g(k)=t+\sum^{i=2}_{i=1} te^{i\bold{k}\bold{a}_{i}}$. Expanding $H_{G}$ around $\vec{K}=\Big(\frac{4\pi}{3a},0\Big)$ and taking into account the intralayer hopping corrections $\delta t_{1},\delta t_{2}$ and $\delta t_{3}$, in $g(k)$, we obtain the following modified graphene Hamiltonian
\begin{align}
    H_{G}=\begin{pmatrix}
    0 & \hbar v(k_{x}-ik_{y})+(A_{x}-iA_{y})\\
    \hbar v(k_{x}+ik_{y})+(A_{x}+iA_{y}) & 0
    \end{pmatrix},
\end{align}
where $v\equiv\frac{-\sqrt{3}}{2a}t$ and
\begin{align}
    &
    A_{x}=\frac{3\beta t}{4}(\frac{\partial u_{x}}{\partial x}-\frac{\partial u_{y}}{\partial y})+O(u^{2}),\\
    &
    A_{y}=-\frac{3\beta t}{4}(\frac{\partial u_{x}}{\partial y}+\frac{\partial u_{y}}{\partial x})+O(u^{2}). \nonumber
\end{align}
\begin{figure}[t!]
    \centering
    \includegraphics[width =7in, height=2.7in]{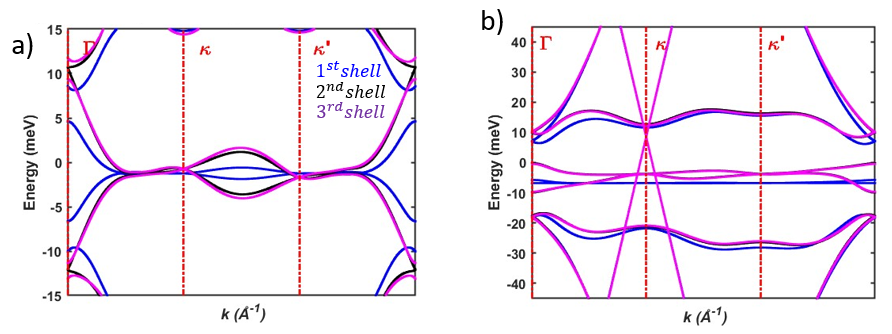}
    \caption{\footnotesize a) Convergence of the bandstructre with increasing number of shells for tBLG at $\theta=0.7^\circ$. b) Convergence of the bandstructure of tTLG for a twist angle of $\theta=1^\circ$. The dimer non-dimer energy difference $\Delta$ is set to zero in both cases to better show the convergence of the bandstructure with an increasing number of shells in the relaxation terms.}
\label{fig:convergence}
\end{figure}
Following the above tight-binding approach, we review the effect that relaxation has in the interlayer tunneling terms following the expansion performed in Ref.\onlinecite{KoshWann},\onlinecite{KN20} and \onlinecite{Xierelax}. Starting from the Bloch state
\begin{align}
    \vert \bold{k},X, l\rangle=\frac{1}{\sqrt{N}}\sum_{R}e^{i\bold{k}\bold{R}}\vert \bold{R}+\bold{u}\rangle,
\end{align}
we firstly define the 2D Fourier transform of the interlayer hopping $T(\vec{r})$ where $\vec{r}$ is the distance between two carbon atoms,
\begin{align}
    t(p)=\frac{1}{S_{0}d_{0}}\int d^{3}r T(\bold{r})e^{-i\bold{p}\bold{r}}.
\end{align}
We therefore obtain the following interlayer tunneling matrix element,
\begin{align}
    &
    \langle \bold{k'},X',2\vert U \vert \bold{k},X,1\rangle=\frac{1}{N}\sum_{R\in l^{1}}\sum_{R'\in l^{2}}e^{-i\bold{k}\bold{R}-i\bold{k'}\bold{R'}}T(\bold{R'}+\bold{u}^{(2)}-\bold{R}-\bold{u}^{(1)}) \\ \nonumber
    &
    =\frac{S_{0}d_{0}}{N(2\pi)^{3}}\sum_{R\in l^{1}}\sum_{R'\in l^{2}}\int d^{3}r T(\bold{r})e^{-i\bold{k}\bold{R}-i\bold{k'}\bold{R'}}e^{i\bold{p}(\bold{R'}+\bold{u_{X'}}^{2}-\bold{R}-\bold{u_{X}}^{1})},
\end{align}
where $S_{0}$ and $d_{0}$ are the unit cell and interlayer distance respectively and $\bold{u}^{1(2)}$ the atomic dispalcement fields of the top and bottom layer respectively. We then expand the exponential function $e^{i\bold{p}\bold{u}^{(l)}}$
and obtain
\begin{align}
    \langle \bold{k}', X', 2\vert U \vert \bold{k}, X, 1\rangle=\sum_{\bold{g,\bold{g}'}}\sum_{n_{1},n_{2},..}\sum_{n'_{1},n'_{2},..} \Gamma^{n'_{1},n'_{2}...}_{n_{1},n_{2},...}(\bold{Q})e^{i(\bold{g}\bold{\tau}^{1}_{X}+\bold{g'}\bold{\tau}^{1}_{X'})}\delta_{\bold{k}+\bold{g}+n_{1}\bold{q}_{1}+n_{2}\bold{q}_{2}+...,\bold{k'}+\bold{g'}-n'_{1}\bold{q}_{1}-n'_{2}\bold{q}_{2}+...},
\end{align}
where $\bold{g}$ are the Moiré reciprocal lattice vectors, $\bold{q}$ are the Fourier modes of $\bold{u}^{(l)}$ and $\bold{Q}=\bold{k}+\bold{g}-n_{1}\bold{q}_{1}-n_{2}\bold{q}_{2}+...$ are the resonant momentum transfer between the $k$ and $k'$ state. Finally, the term $\Gamma$ is defined as,
\begin{align}
\Gamma^{n'_{1},n'_{2}...}_{n_{1},n_{2},...}=t_{||}(\bold{Q},d_{0})\Big[\frac{-i\bold{Q}\bold{u}^{1}_{X\bold{q}_{1}}}{n_{1}!}\Big]^{n_{1}}\Big[\frac{-i\bold{Q}\bold{u}^{1}_{X\bold{q}_{2}}}{n_{2}!}\Big]^{n_{1}}...\times\Big[\frac{-i\bold{Q}\bold{u}^{2}_{X'\bold{q}_{1}}}{n'_{1}!}\Big]^{n'_{1}}\Big[\frac{-i\bold{Q}\bold{u}^{2}_{X'\bold{q}_{2}}}{n'_{2}!}\Big]^{n'_{2}}...,\label{eqn:Gamma}
\end{align}
where $t_{||}(Q,d_{0})$ is the $Q$ Fourier mode of $T(r)$ at a fixed interlayer distance $d_{0}$ . Eq.\ref{eqn:Gamma} clearly shows the very quick decay of the corrections in the interlayer tunneling terms with an increasing number of harmonics. Using the strain field formula in Eq.(\ref{eqn:NonlinU}) and Eq.(\ref{eqn:NonlinUy}) we obtained interlayer tunnelings corrections $\delta\omega/\omega$ ranging from $0.3$ to $0.5\%$ by simply replacing $\vec{r}$ by $\vec{r}+\vec{u}$ in Eq.(\ref{eqn:BM}), permitting us to safely ignore $\delta \omega$ in our continuum Hamiltonian. 

\section{Two-band model of fully relaxed \MakeLowercase{t}BLG around the K-point}\label{app:twoband}

To find the renormalized Fermi velocity at sufficiently large twist angles $\theta$, we will first formulate a low-energy 2-band model of tBLG around the $K$-point accounting for atomic relaxation. To commence, we start by truncating the BM Hamiltonian into a simpler $8\times 8$ matrix Hamiltonian which only considers the resonant tunneling between the bottom layer Dirac cone and its three nearest top layer counterparts. This approximate model was demonstrated in Ref.\onlinecite{BM11} to provide a very accurate prediction for the first magic angle in tBLG. As previously discussed in Section.\ref{section:cont}, lattice relaxation adds a three-fold symmetric pseudogauge field $\mathcal{A}$ connecting Dirac cones separated by integer multiples of the Moiré reciprocal lattice vectors $\vec{G}_{1}=\frac{4\pi}{\sqrt{3}L}(1,0),\vec{G}_{2}=\frac{4\pi}{\sqrt{3}L}(-1/2,\sqrt{3}/2)$ and $\vec{G}_{3}=\frac{4\pi}{\sqrt{3}L}(-1/2,-\sqrt{3}/2)$. Therefore, the tripod Hamiltonian of fully relaxed tBLG written in the basis $(\Psi^{A1},\Psi^{B1},\Psi^{B2},\Psi^{B3})$ designating the four Dirac cones depicted in Fig.\ref{fig:magic} a) becomes 
\begin{align}\label{eqn:tripintro}
    H\Psi=\begin{pmatrix}
   v_{F}\bold{k}\vec{\sigma} & \mathcal{T}^{1} & \mathcal{T}^{2} & \mathcal{T}^{3} \\
   \mathcal{T}^{1\dag} & v_{F}(\bold{k}-\bold{q}_{1})\vec{\sigma} & \mathcal{A}^{3}\vec{\sigma} & \mathcal{A}^{2}\vec{\sigma} \\
    \mathcal{T}^{2\dag} & \mathcal{A}^{3}\vec{\sigma} & v_{F}(\bold{k}-\bold{q}_{2})\vec{\sigma} & \mathcal{A}^{1}\vec{\sigma} \\
    \mathcal{T}^{3\dag} & \mathcal{A}^{2}\vec{\sigma} & \mathcal{A}^{1}\vec{\sigma} & v_{F}(\bold{k}-\bold{q}_{3})\vec{\sigma}
    \end{pmatrix}\begin{pmatrix}
   \Psi^{A1} \\
   \Psi^{B1} \\
    \Psi^{B2} \\
    \Psi^{B3}\end{pmatrix}=E\begin{pmatrix}
   \Psi^{A1} \\
   \Psi^{B1} \\
    \Psi^{B2} \\
    \Psi^{B3}
    \end{pmatrix},
\end{align}
where the interlayer tunneling matrices $\mathcal{T}^{j}$ and the gauge field $\mathcal{A}^{j}$ were previously defined in Eq.\ref{eqn:matrixdef} of Section.\ref{section:cont} while the vectors $\vec{q}_{1},\vec{q}_{2},\vec{q}_{3}$ are $\Delta K(0,1),\Delta K(-\sqrt{3}/2,-1/2)$ and $\Delta K(\sqrt{3}/2,-1/2)$ respectively. The $8\times 8$ Hamiltonian in Eq.\ref{eqn:tripintro} is then subdivided into two square quadrants $Q_{l}$ and $Q_{h}$,
\begin{align}
    Q_{l}=v_{F}\bold{k}\vec{\sigma},\quad Q_{h}= \begin{pmatrix}
    v_{F}(\bold{k}-\bold{q}_{1})\vec{\sigma} & \mathcal{A}^{3}\vec{\sigma} & \mathcal{A}^{2}\vec{\sigma} \\
     \mathcal{A}^{3}\vec{\sigma} & v_{F}(\bold{k}-\bold{q}_{2})\vec{\sigma} & \mathcal{A}^{1}\vec{\sigma} \\
     \mathcal{A}^{2}\vec{\sigma} & \mathcal{A}^{1}\vec{\sigma} & v_{F}(\bold{k}-\bold{q}_{3})\vec{\sigma}
    \end{pmatrix},
\end{align}
which are connected by the interlayer tunnelling terms $\mathcal{T}^{j}$. The matrices $Q_{l}$ and $Q_{h}$ can be identified as the low and the high-energy states of the tripod Hamiltonian. Considering the diagonal $v_{F}\bold{k}\vec{\sigma}$ terms as a perturbation in $Q_{h}$, we perform a unitary transformation $\hat{U}$ in $Q_{h}$ which diagonalises the $Q_{h}$ matrix,
\begin{align}
    & 
    \Tilde{\bold{Q}}_{h}=\hat{U}^{-1}Q_{h}\hat{U}=\begin{pmatrix}
     (A+v_{F}\Delta K)\mathbb{I}_{2} & 0 & 0 \\
     0 & -(A+v_{F}\Delta K)\mathbb{I}_{2} & 0 \\
     0 & 0 & (-2A+v_{F}\Delta K)\sigma_{z}
    \end{pmatrix}, \text{where\quad} \\ \nonumber  
    & U=\begin{pmatrix}
  -\frac{1}{2} e^{-\frac{i\pi}{4}} & \frac{e^{-\frac{i\pi}{4}}}{2 \sqrt{3}} & \frac{1}{2} e^{-\frac{i\pi}{4}} & -\frac{e^{-\frac{i\pi}{4}}}{2 \sqrt{3}} & \frac{e^{-\frac{i\pi}{4}}}{\sqrt{6}} & -\frac{e^{-\frac{i \pi}{4}}}{\sqrt{6}} \\
   0 & \frac{e^{-\frac{i\pi}{12}}}{\sqrt{3}} & 0 & \frac{e^{-\frac{i\pi}{12}}}{\sqrt{3}} & -\frac{e^{-\frac{i\pi}{12}}}{\sqrt{6}} & -\frac{e^{-\frac{i\pi}{12}}}{\sqrt{6}} \\
  0 & \frac{e^{\frac{i\pi }{12}}}{\sqrt{3}} & 0 & -\frac{e^{\frac{i \pi }{12}}}{\sqrt{3}} & -\frac{e^{\frac{i \pi }{12}}}{\sqrt{6}} & \frac{e^{\frac{i \pi }{12}}}{\sqrt{6}} \\
  \frac{1}{2} e^{-\frac{5i\pi}{12}} & \frac{e^{-\frac{5i\pi}{12}}}{2 \sqrt{3}} & \frac{1}{2} e^{-\frac{5i\pi}{12}} & \frac{e^{-\frac{5i\pi}{12}}}{2\sqrt{3}} & \frac{e^{-\frac{5i\pi}{12}}}{\sqrt{6}} & \frac{e^{-\frac{5i\pi}{12}}}{\sqrt{6}} \\
  \frac{1}{2} e^{\frac{5i\pi }{12}} & \frac{e^{\frac{5i\pi }{12}}}{2 \sqrt{3}} & -\frac{1}{2} e^{\frac{5i\pi }{12}} & -\frac{e^{\frac{5i\pi }{12}}}{2 \sqrt{3}} & \frac{e^{\frac{5i\pi }{12}}}{\sqrt{6}} & -\frac{e^{\frac{5i\pi }{12}}}{\sqrt{6}} \\
\end{pmatrix}.
\end{align}
Regarding $Q_{l}$, we projected this $2\times2$ Hamiltonian in the orthogonal basis defined by its two eigenvectors $\Psi_{c}=\frac{1}{\sqrt{2}}\begin{pmatrix}
     e^{-i\theta_{k}/2} \\ e^{i\theta_{k}/2}
\end{pmatrix}$ and $\Psi_{v}=\frac{1}{\sqrt{2}}\begin{pmatrix}
     e^{-i\theta_{k}/2} \\ -e^{i\theta_{k}/2}
\end{pmatrix}$, where $\theta_{k}\equiv\arctan(k_{y}/k_{x})$. Hence, the diagonal matrix emanating from this projection has the form,
\begin{align}
     \Tilde{\bold{Q}}_{l}=O^{-1}_{\theta_{k}}Q_{l}O_{\theta_{k}}=\begin{pmatrix}
     v_{F}k & 0 \\ 0 & -v_{F}k \end{pmatrix} \text{where \quad} O_{\theta_{k}}=\frac{1}{\sqrt{2}}\begin{pmatrix}
     e^{-i\theta_{k}/2} & e^{-i\theta_{k}/2} \\ e^{i\theta_{k}/2} & -e^{i\theta_{k}/2} \end{pmatrix}.
\end{align}
For simplicity, we will fix the momentum orientation $\theta_{k}$ in the $\kappa-\kappa'$ direction ($\theta_{k}=\pi/2$), neglecting the trigonal symmetry around the Dirac point. From the matrices $U$ and $O_{\theta_{k}}$, we construct a projection operator $\mathcal{P}$ defined as
\begin{align}\label{eqn:Pmatrix}
\mathcal{P}\equiv\begin{pmatrix}
     O_{\theta_{k}} & \tilde{0} \\ \tilde{0} & U\end{pmatrix},
\end{align}
where $\tilde{0}$ is a $2\times 6$ zero matrix. Considering the limit of $|\vec{k}|\longrightarrow 0$ in $\Tilde{\bold{Q}}_{l}$, we split $H_{trip}$ into two parts, $H_{0}$ and $H'$ labeling the bare and the perturbative part of the tripod Hamiltonian. Within $H_{0}$ we have the diagonal matrices $\Tilde{\bold{Q}}_{l}$ in the limit of $\bold{k}\longrightarrow0$ as well as $\Tilde{\bold{Q}}_{h}$ and the projected interlayer tunneling terms which become $\mathcal{T}^{j}\longrightarrow O_{\pi/2}^{-1}\mathcal{T}^{j}U$,
\begin{align}
    &
    H_{0}=\begin{pmatrix}
        H^{t}_{0} & \bold{0} \\
        \bold{0}^{T} & H^{b}_{0}
    \end{pmatrix}, \\ \nonumber
\end{align}
where $\bold{0}$ is a $6\times 2$ zero matrix independent of $\theta_{k}$, $H^{b}_{0}=(-2A+v_{F}\Delta K)\sigma_{z}$ and 
\begin{align}\label{eqn:Bare}
    H^{t}=\left(
\begin{array}{cccccc}
0 & 0 & \frac{\sqrt{3}\omega_{0}-3\omega_{1}}{2 \sqrt{2}} & \frac{1}{2} \sqrt{\frac{3}{2}} \left(\sqrt{3} \omega_{0}+\omega_{1}\right) & -\frac{i \left(3\omega_{0}-\sqrt{3}\omega_{1}\right)}{2 \sqrt{2}} & \frac{1}{2} i \sqrt{\frac{3}{2}} \left(\omega_{0}+\sqrt{3}\omega_{1}\right) \\
 0 & 0 & -\frac{i \left(3\omega_{0}+\sqrt{3}\omega_{1}\right)}{2 \sqrt{2}} & \frac{1}{2} i \sqrt{\frac{3}{2}} \left(\omega_{0}-\sqrt{3}\omega_{1}\right) & \frac{\sqrt{3}\omega_{0}+3\omega_{1}}{2 \sqrt{2}} & \frac{1}{4} \left(3 \sqrt{2}\omega_{0}-\sqrt{6}\omega_{1}\right) \\
\frac{\sqrt{3}\omega_{0}-3\omega_{1}}{2 \sqrt{2}} & \frac{i\left(3\omega_{0}+\sqrt{3}\omega_{1}\right)}{2 \sqrt{2}} & A+v_{F}\Delta K & 0 & 0 & 0  \\
\frac{3\omega_{0}+\sqrt{3}\omega_{1}}{2 \sqrt{2}} & -\frac{i \left(\sqrt{3}\omega_{0}-3\omega_{1}\right)}{2 \sqrt{2}} & 0 & A+v_{F}\Delta K & 0 & 0 \\
\frac{i \left(3\omega_{0}-\sqrt{3}\omega_{1}\right)}{2 \sqrt{2}} & \frac{\sqrt{3}\omega_{0}+3\omega_{0}}{2 \sqrt{2}} & 0 & 0 & -(A+v_{F}\Delta K) & 0 \\
 -\frac{i \left(\sqrt{3}\omega_{0}+3\omega_{1}\right)}{2 \sqrt{2}} & \frac{3 \omega_{0}-\sqrt{3}\omega_{1}}{2 \sqrt{2}} & 0 & 0 & 0 & -(A+v\Delta K) \\ 
\end{array}
\right).
\end{align}
Since there is no coupling between the zero-energy states in $H^{t}_{0}$ and the high-energy levels in $H^{b}_{0}$, we will neglect $H^{b}_{0}$ in the unperturbed Hamiltonian such that $H_{0}=H^{t}_{0}$. \\
\indent The perturbative term $H'$ consist of the diagonal $v_{F}\bold{k}\vec{\sigma}$ terms projected in the orthogonal basis $\mathcal{P}$ defined in Eq.\ref{eqn:Pmatrix}. Akin to $H_{0}$, the coupling terms between the $A+v_{F}\Delta K$ and the $(-2A+v_{F}\Delta K)$ states are neglected as they would contribute quadratically to the low-energy dispersion of $H^{t}_{0}$. Therefore, the perturbative $H'$ term becomes  
\begin{align}\label{eqn:pert}
    H'=v_{F}\left(
\begin{array}{cccccc}
 k & 0 & 0 & 0 & 0 & 0 \\
 0 & -k & 0 & 0 & 0 & 0 \\
 0 & 0 & -\frac{k}{4} & \frac{\sqrt{3}k}{4} & -\frac{1}{4} i \sqrt{3}k & -\frac{1}{4}(ik) \\
 0 & 0 & \frac{\sqrt{3}k}{4} & \frac{k}{4} & -\frac{1}{4}(ik) & \frac{1}{4}i\sqrt{3}k \\
 0 & 0 & \frac{1}{4}i\sqrt{3}k & \frac{i k}{4} & \frac{k}{4} & -\frac{1}{4} \left(\sqrt{3}k\right) \\
 0 & 0 & \frac{i k}{4} & -\frac{1}{4} i \sqrt{3} k & -\frac{1}{4} \left(\sqrt{3} k\right) & -\frac{k}{4} \\
\end{array}
\right).
\end{align}
The doubly degenerate zero-energy eigenstates of $H_{0}$ were obtained by diagonalizing $H_{0}$,
\begin{align}
    \Psi^{c}=\frac{1}{\Big(2+\frac{2(A+v_{F}\Delta K)^{2}}{3(\omega^{2}_{0}+\omega^{2}_{1})}\Big)}
\begin{pmatrix}
 \frac{i (A+v_{F}\Delta K) \left(\omega^{2}_{0}+2 \sqrt{3}\omega_{0}\omega_{1}+3\omega_{1}^2\right)}{\sqrt{2}(\sqrt{3}\omega_{0}+3\omega_{1})(\omega^{2}_{0}+\omega_{1}^2} \\
 \frac{(A+v_{F}\Delta K) \left(3\omega_{0}-\sqrt{3}\omega_{1}\right)}{3 \sqrt{2} \left(\omega^{2}_{0}+\omega^{2}_{1}\right)}\\
 -\frac{i \left(\omega^{2}_{0}-\omega^{2}_{1}\right)}{\omega^{2}_{0}+\omega^{2}_{1}}\\
 -\frac{2 i \omega_{0}\omega_{1}}{\omega^{2}_{0}+\omega^{2}_{1}} \\
 0 \\
 1 \\
\end{pmatrix}, \Psi^{v}=
\frac{1}{\Big(2+\frac{2(A+v_{F}\Delta K)^{2}}{3(\omega^{2}_{0}+\omega^{2}_{1})}\Big)}\begin{pmatrix}
 -\frac{i (A+v_{F}\Delta K) \left(3\omega_{0}-\sqrt{3}\omega_{1}\right)}{3 \sqrt{2} \left(\omega^{2}_{0}+\omega^{2}_{1}\right)} \\
 \frac{(A+v_{F}\Delta K) \left(\sqrt{3}\omega_{0}+3\omega_{1}\right)}{3 \sqrt{2} \left(\omega^{2}_{0}+\omega^{2}_{1}\right)} \\
-\frac{2 i\omega_{0}\omega_{1}}{\omega^{2}_{0}+\omega^{1}_{0}} \\
 \frac{i \left(\omega^{2}_{0}-\omega^{2}_{1}\right)}{\omega^{2}_{0}+\omega^{2}_{1}} \\
  1 \\
  0 \\
\end{pmatrix}.
\end{align}
This new set of eigenstates permits to construct a 2-band model projecting $H'$ into the orthogonal basis ($\Psi^{c}$,$\Psi^{v}$),
\begin{align}
    &
    H_{eff}=\begin{pmatrix}
    \langle \Psi^{c} |H'| \Psi^{c} \rangle & \langle \Psi^{c} |H'| \Psi^{v}\rangle \\
    \langle \Psi^{v} |H'| \Psi^{c} \rangle & \langle \Psi^{v} |H'| \Psi^{v} \rangle
    \end{pmatrix}= \\ \nonumber
    &
    v_{F}\begin{pmatrix} \frac{3 k(-\omega^{4}_{0}+10\omega_{0}^2 \omega_{1}^2+8\sqrt{3}\omega_{0}\omega_{1}^{3}+3\omega_{1}^{4})((A+v_{F}\Delta K)^{2}-3\omega_{1}^2)}{2 (\sqrt{3}\omega_{0}+3\omega_{1})^{2} (\omega_{0}^{2}+\omega_{1}^{2})((A+v_{F}\Delta K)^{2}+3 (\omega^{2}_{0}+\omega^{2}_{1}))} & \frac{k(-3\omega_{0}^{3}-5\sqrt{3}\omega_{0}^{2}\omega_{1}-3 \omega_{0}\omega^{2}_{1}+3\sqrt{3}\omega_{1}^3) ((A+v_{F}\Delta K)^{2}-3\omega^{2}_{1})}{2(\sqrt{3}\omega_{0}+3\omega_{1})(\omega_{0}^2+\omega_{1}^2)((A+v_{F}\Delta K)^2+3(\omega^{2}_{0}+\omega^{2}_{1}))} \\
 \frac{k(-3\omega_{0}^{3}-5\sqrt{3}\omega_{0}^{2}\omega_{1}-3\omega_{0} \omega_{1}^2+3\sqrt{3}\omega_{1}^3)((A+v_{F}\Delta K)^{2}-3 \omega_{1}^2)}{2(\sqrt{3} \omega_{0}+3\omega_{1})(\omega_{0}^2+\omega_{1}^2) ((A+v\Delta K)^{2}+3(\omega_{0}^{2}+\omega_{1}^2))} & \frac{k (\omega_{0}^2-2 \sqrt{3}\omega_{0}\omega_{1}-\omega_{1}^2)(\alpha'^2-3\omega_{1}^2)}{2(\omega_{0}^2+\omega_{1}^2)(\alpha'^{2}+3 (\omega^{2}_{0}+\omega^{2}_{1}))}. \\
\end{pmatrix}.
\end{align}
Diagonalizing $H_{eff}$, the Fermi velocity was obtained for three different situations. Firstly, when $\omega_{1}=\omega_{0}$, secondly for $\omega_{1}\neq\omega_{0}$ and thirdly for $\omega_{0}=0$,
\begin{align}\label{eqn:fermivel}
    &
    v_{\omega_{0}=\omega_{1}}=v_{F}\frac{(1-3\alpha'^{2})}{(1+6\alpha'^{2})}, \\ \nonumber
    &
    v_{\omega_{0}\neq\omega_{1}}=v_{F}\frac{3(-(A+v_{F}\Delta K)^{2}+3\omega^{2}_{1})\sqrt{(\omega^{2}_{0}+\omega^{2}_{1})^{2}(\omega^{4}_{0}+4\sqrt{3}\omega^{3}_{0}\omega_{1}+18\omega^{2}_{0}\omega^{2}_{1}+12\sqrt{3}\omega_{0}\omega^{3}_{1}+9\omega^{4}_{1})}}{(\sqrt{3}\omega_{0}+3\omega_{1})(\omega^{2}_{0}+\omega^{2}_{1})((A+v_{F}\Delta K)^{2}+3(\omega^{2}_{0}+\omega^{2}_{1})}, \\  \nonumber
    &
    v_{\omega_{0}=0}=v_{F}\frac{(1-3\alpha'^{2})}{(1+3\alpha'^{2})}.  \\ \nonumber
\end{align}
where $\alpha'=\omega_{1}/(A+v_{F}\Delta K)$.We noted that in the limit of no relaxation $(A\longrightarrow 0)$, the expressions for $v_{chiral}$ and $v_{\omega_{1}=\omega_{0}}$ coincided with the ones reported in Ref.\onlinecite{BM11} and \onlinecite{VafekRelax} respectively. Furthermore, from Eq.\ref{eqn:fermivel} and Eq.\ref{eqn:Bare} it is transparent that the effect of atomic relaxation is no other than a simple replacement of $v_{F}\Delta K$ by $v_{F}\Delta K+A$, effectively increasing the twist angle in the AA-sites.

\end{widetext}

\end{document}